\documentclass[12pt]{JHEP} % 10pt is ignored!
\usepackage{epsfig}

\newcommand{\Tr}{{\rm Tr}\ }

\newcommand{\W}{{\cal W}}
\newcommand{\N}{{\cal N}}
\newcommand{\Path}{\mbox{P}}

\newcommand{\feyn}[1]{\raisebox{-0.4cm}{\epsfxsize=1cm\epsfbox{#1.eps}}}
\newcommand{\be}{\begin{equation}}
\newcommand{\ee}{\end{equation}}
\newcommand{\bea}{\begin{eqnarray}}
\newcommand{\eea}{\end{eqnarray}}
\newcommand{\ft}[2]{{\textstyle\frac{#1}{#2}}\,}
\newcommand\nn{\nonumber}
\newcommand\Wurzel{1-\sqrt{\ft{(R_1+R_2)^2+h^2}{(R_1-R_2)^2 + h^2}}}
\newcommand{\eqn}[1]{(\ref{#1})}
\newcommand{\xd}[1]{\dot x_{#1}}
\newcommand{\yd}[1]{\dot y_{#1}}
\newcommand{\sign}{ \mbox{sgn} }

\title{Two Loops to Two Loops in ${\cal N}=4$ Supersymmetric 
Yang-Mills Theory}

\author{Jan Plefka and Matthias Staudacher \\
Albert-Einstein-Institut, Max-Planck-Institut
f\"ur Gravitationsphysik\\Am M\"uhlenberg 1, D-14476 Golm, Germany\\
Email: \email{plefka,matthias@aei-potsdam.mpg.de}}

\preprint{\hepth{0108182}\\
AEI-2001-106\\}  

\abstract{We present a full two-loop ${\cal O}(g^6)$ perturbative
field theoretic calculation of the expectation value of two circular 
Maldacena-Wilson loops in $D=4$ ${\cal N}=4$ supersymmetric $U(N)$ 
gauge theory. It is demonstrated that, after taking into account
very subtle cancellations of bulk and boundary divergences, the 
result is completely finite without any renormalization. As opposed 
to previous lower order calculations existing in the literature, 
internal vertex diagrams no longer cancel identically and lead to 
subleading corrections to the dominant ladder diagrams. 
Taking limits, we proceed to extract the two-loop static 
potential corresponding to two infinite anti-parallel lines.
Our result gives some evidence that the existing strong-coupling 
calculations using the AdS/CFT conjecture might sum up the 
full set of large $N$ planar Feynman diagrams.
}


\begin{document}

\section{Introduction and Conclusions}
Interest in 't Hooft's 1973 proposal \cite{'tHooft:1974jz}
to consider the limit $N \rightarrow \infty$ of $U(N)$ quantum gauge theory
such that $\lambda=N g^2$ stays finite
is not waning even after nearly thirty years of, to date,
futile attempts to analytically perform the sum over the resulting 
planar Feynman diagrams. In the original work it was already conjectured that
an indirect approach using string theory might be more likely to
succeed. The first-ever concrete suggestion for formulating such a
string theory ``dual'' to the very special four-dimensional gauge
theory invariant under ${\cal N}=4$ supersymmetries
has been made by Maldacena \cite{Maldacena:1998re}. In its most modest
form the conjecture holds that the limit of strong `t~Hooft coupling
$\lambda \rightarrow \infty$ in the $N=\infty$ gauge theory is reproduced by
the low energy supergravity limit of IIB string theory in the
background AdS$_5 \times S^5$. This proposal does lead to very 
explicit, analytic predictions for the large $N$ and $\lambda$ limit
of various observables in the gauge theory. It is usually believed
that instanton effects are inessential at $N=\infty$; therefore
the predictions should coincide with the two-step procedure of 
first calculating the observables in question to all orders in planar 
perturbation theory and, barring large $N$ phase transitions,
subsequently taking the large $\lambda$ limit 
of the resulting sum. Our inability, however, of actually carrying through
this program constituted the chief motivation for seeking a dual formulation
in the first place. In short, we are in the frustrating situation
of being presented with a proposed solution of a notorious problem without
being able to prove that the solution is indeed correct!

Recently, there has been some interesting progress towards making
much more direct contact between the Euclidean ${\cal N}=4$ gauge theory and 
its suggested  supergravity dual. In \cite{Maldacena:1998im} 
Maldacena introduced a non-local loop operator, which differs from
the usual Wilson loop in a subtle way: In addition to the gauge
field the six scalars of the model are also coupled to the loop $C$. The
coupling is such that the new operator $\W[C]$, which we will
call a Maldacena-Wilson loop operator, is no longer unitary and its
magnitude is thus no longer bounded by one (see eq.(\ref{maldaloop}) for a
precise definition). A priori it is therefore not even clear whether
the expectation value of this operator exists. AdS/CFT, for
various types of contours $C$, leads to some very explicit results
\cite{Maldacena:1998im},\cite{Berenstein:1999ij}, see also
\cite{Aharony:2000ti}. A straight infinite
line e.g.~is a BPS object and should satisfy
\be
\langle \W[C] \rangle =1, \qquad  C={\rm straight} \; {\rm line}.
\label{straight}
\ee
Furthermore, in the 
double limit $N \rightarrow \infty$, $\lambda=N g^2 \rightarrow \infty$   
one finds for a  circular contour, independently of the radius,
\be 
\langle \W[C] \rangle = e^{\sqrt{\lambda}},
\qquad C={\rm circle}.
\label{circle}
\ee
Finally for a rectangular Maldacena-Wilson loop of sides $T \times L$ the 
prediction reads, for $N=\infty$, $\lambda \rightarrow
\infty$ and $T \rightarrow \infty$,
\be
\langle \W[C] \rangle = \exp \Bigg( 
{4 \pi^2 \over \Gamma^4(1/4) }\sqrt{\lambda} {T \over L} \Bigg).
\qquad C={\rm rectangle},
\label{rectangle}
\ee  
which may be used to define the so-called static potential
$V(L)=-1/T \log \langle \W[C] \rangle$.

In \cite{Erickson:2000qv},\cite{Erickson:2000af}
the authors pioneered the weak coupling, perturbative analysis of 
Malda\-cena-Wilson loops. They found that up to
one-loop order (i.e.~${\cal O}(g^4)$) all Feynman diagrams cancel for a 
straight line,
as is required if eq.(\ref{straight}) is to hold in the gauge theory.
Furthermore, for a circle and a rectangle (with $T \rightarrow
\infty$, i.e. infinite anti-parallel lines) 
all non-ladder diagrams (i.e.~containing internal interaction
vertices) cancel. The authors went on to sum up ladder diagrams
to all orders for the circle and the anti-parallel
lines. Interestingly, for $N=\infty$ (planar ladders) and for the circle 
they found {\it exactly} eq.(\ref{circle}), while the anti-parallel
lines yielded something very similar to eq.(\ref{rectangle}):
\be
\langle \W[C] \rangle_{{\rm ladder}~{\rm approx.}} = \exp \Bigg( 
{1 \over \pi }\sqrt{\lambda} {T \over L} \Bigg).
\qquad C={\rm rectangle}.
\label{ladderpotential}
\ee  
Closer inspection of these calculations reveals a number of
interesting features:

\noindent$\bullet$ The celebrated $\sqrt{\lambda}$ strong 
coupling ``screening''
behavior of eqs.(\ref{circle}),(\ref{rectangle}) is seen to be
a large $N$ artifact: If one looks at the ladder approximation at
{\it finite} $N$ one easily proves that the behavior is always
$\sim \exp g^2 \sim \exp{\lambda/N}$. 
Incidentally, the ladder approximation has been
argued to be exact for a circle even at finite $N$ and $\lambda$
\cite{Drukker:2000rr}. If true, the result for, say, $SU(2)$ would
read
\be
\langle \W[C] \rangle_{{\rm SU}(2)}=
(1+{1 \over 8} g^2)~e^{{1 \over 16} g^2}
\qquad C={\rm circle}.
\label{su2}
\ee

\noindent$\bullet$ Individual non-ladder diagrams are divergent.
After adding all interactive diagrams the divergences cancel.
It is particularly intriguing that some divergent contact interactions 
at the boundaries of the loops (which are present in an ordinary unitary 
Wilson loop, even in ${\cal N} =4$ theory) are precisely 
canceled by a bulk divergence in the supersymmetric self-energy.
This demonstrates that the Maldacena-Wilson loop might turn out to be a 
field theoretic observable that is finite to all orders in 
perturbation theory! A proof of this is, however, currently lacking.

\noindent$\bullet$ For the circle the ladder approximation reproduces
the AdS/CFT strong coupling result, which is consistent with it 
being exact. For the anti-parallel lines we notice by comparing
eq.(\ref{rectangle}) and eq.(\ref{ladderpotential}) that the result 
is qualitatively, but not quantitatively correct. Turning this
around we see that AdS/CFT {\it predicts} that non-ladder diagrams
must contribute to the static potential: The cancellation mechanism of 
internal vertex diagrams at ${\cal O}(g^4)$ found in 
\cite{Erickson:2000qv},\cite{Erickson:2000af} should not
extend to all orders in perturbation theory, for in that case 
the AdS/CFT correspondence, in its current form, would be proven wrong.

In the present paper we are aiming at extending and deepening the results 
of \cite{Erickson:2000qv},\cite{Erickson:2000af}. Clearly one should
perform a two-loop calculation in order to investigate
whether internal vertex diagrams begin to contribute, as well as
to further test the finiteness of the Maldacena-Wilson loop. Unfortunately
the number of diagrams relevant to the so far mentioned contours
is discouragingly large. We therefore found it convenient to study
a slightly different situation which nevertheless allows to gain
new insight: We will consider two closed, axisymmetric, parallel circles.
As will be shown in the following section a large number of
diagrams are zero for group theoretic reasons and a two-loop calculation
becomes feasible (see the figure in section 2 for the relevant diagrams).
The configuration of two circular loops has been studied before in the
literature. Strong-coupling supergravity computations were done
in \cite{Zarembo:1999bu}, and it was found, notably, that the limit of
very large, nearby circles exactly reproduces the anti-parallel lines 
potential of eq.(\ref{rectangle}).
Furthermore, the sum over all ladder diagrams for this 
situation was obtained in \cite{Zarembo:2001jp}, reproducing, in the
anti-parallel lines limit, the result of eq.(\ref{ladderpotential}).
Hence from the strong-coupling perspective our scenario of two parallel,
axisymmetric circles does not seem to differ from the anti-parallel lines
situation in the static potential limit.

Despite the considerable reduction of the number of diagrams mentioned
above our two-loop 
perturbative calculation is  quite lengthy and will be presented
below. We decided to detail some of the techniques employed since they
might be found useful for further investigations. 
 
For our geometry we are able to exhibit the complete finiteness of the
Maldacena-Wilson loop up to ${\cal O}(g^6)$. Some interactive
diagrams (see the X- and H-graphs in sections 3.4 and 3.5) are 
superficially divergent due to contact interactions at the circular boundaries.
However, due to the special nature of the loop operator
the divergent contributions cancel between gauge and scalar degrees of
freedom. An even more intricate cancellation takes place between a
divergent ``bulk'' self-energy contribution (section 3.2) 
and a divergent ``boundary'' contact interaction 
(the IY-graph in section 3.3), repeating a phenomenon first 
discovered in \cite{Erickson:2000af}. We find this perturbative finiteness of
the Maldacena-Wilson observable quite remarkable; it certainly deserves
a deeper understanding.

Adding the finite contributions of all non-ladder diagrams we establish 
that, to two-loop order, {\it they no longer cancel}. For finite geometry,
this requires some numerical analysis in the final steps.
Taking, in section 4, the limit of very large, near-by circles 
we analytically complete the computation and extract the
static potential limit corresponding to infinite anti-parallel 
lines\footnote{Due to the circular boundary conditions our static potential
result differs to ${\cal O}(g^4)$ from the perturbative calculations
of \cite{Erickson:2000qv},\cite{Erickson:2000af}. As mentioned above,
this difference is a weak coupling feature that is expected to
go away at strong coupling.}. It is then proved that the 
vertex-diagrams contribute in a weak, but non-vanishing fashion.
We will end by discussing whether these weak, non-zero corrections to
the ladder approximation might be of help in explaining the discrepancy
between eqs.\eqn{rectangle} and \eqn{ladderpotential}.

In an outlook (section 5) we collect some of the exciting unsolved
mysteries surrounding Maldacena-Wilson loops.

\section{The Graphs and the Computation to Order $g^2$ and $g^4$}

An interesting modification of the standard (Euclidean) Wilson loop operator,
appropriate for ${\cal N}=4$ gauge theory,
was introduced in \cite{Maldacena:1998im}. Its transformation
properties under supersymmetry were elucidated in 
\cite{Drukker:1999zq}.
It couples not only to the gauge potentials $A_\mu(x)$, 
but also to the six scalar fields $\Phi_I(x)$
of $\N=4$ SYM theory
\be
\W[C]=\Tr \Path \exp \Bigl [\oint_C d\tau(iA_\mu(x)\dot x^\mu + \Phi_I(x)\theta^I |\dot x|)
\Bigr ]
\label{maldaloop}
\ee 
Here $\theta^I$ is a point on the unit five-sphere, 
i.e.~$\theta^I\theta^I=1$, and
$x^\mu(\tau)$ parameterizes the curve $C$. The gauge field and the scalars are
in the fundamental representation of $U(N)$ with generators 
$T^a$ ($a=0,\ldots,N^2-1$),
which we normalize according to $\Tr T^a\, T^b=\ft 1 2 \delta^{ab}$ and 
which obey the $U(N)$ algebra
\be
[T^a,T^b]=if^{abc}T^c
\ee
Note that $\Tr T^a= \sqrt{\ft N2}\delta^{a0}$, $f^{0bc}=0$ as well as 
$f^{abc}f^{abc}=N(N^2-1)$.
Due to the absence of an $i$ in front of the scalar fields $\Phi_I$
in eq.(\ref{maldaloop}) the Maldacena-Wilson operator differs from an 
ordinary Wilson operator in an important way: It is no longer a pure
phase factor, and therefore not bounded in field space.

We employ the Euclidean action of ${\cal N}=4$ supersymmetric Yang-Mills theory
\bea
S&=&  \ft {1}{2g^2}\int d^4x\Bigl [ \ft 1 2 (F^a_{\mu\nu})^2+ ( D_\mu\Phi_I^a)^2
+i\bar\psi^a\gamma^\mu D_\mu\psi^a
+ i f^{abc}\bar\psi^a\Gamma^I\Phi_I^b\psi^c \nn\\&&\qquad\qquad
+\ft 12 (f^{abc}\Phi_I^b\Phi_J^c)^2+ \partial_\mu\bar c ^a\, D_\mu c^a+
(\partial_\mu A_\mu^a)^2 \Bigr ]
\label{action}
\eea
in the Feynman gauge with ghosts $c^a$ and $\bar c^a$. 
Here $D_\mu(\cdot)^a=\partial_\mu(\cdot)^a+f^{abc}A_\mu^b(\cdot)^c$, $\psi^a$ is a sixteen component Majorana spinor and $(\gamma^\mu,\Gamma^I)$
are ten $16\times 16$ Dirac matrices stemming from the reduction of the
ten dimensional model.

%\begin{figure}[t]
\FIGURE[t]{
%\begin{center}
\begin{tabular}{|l|l|l|}
\hline% & &\cr
 & Ladder-Graphs &  Non-Ladder-Graphs\cr
\hline & &\cr
$g^2$:
& \feyn{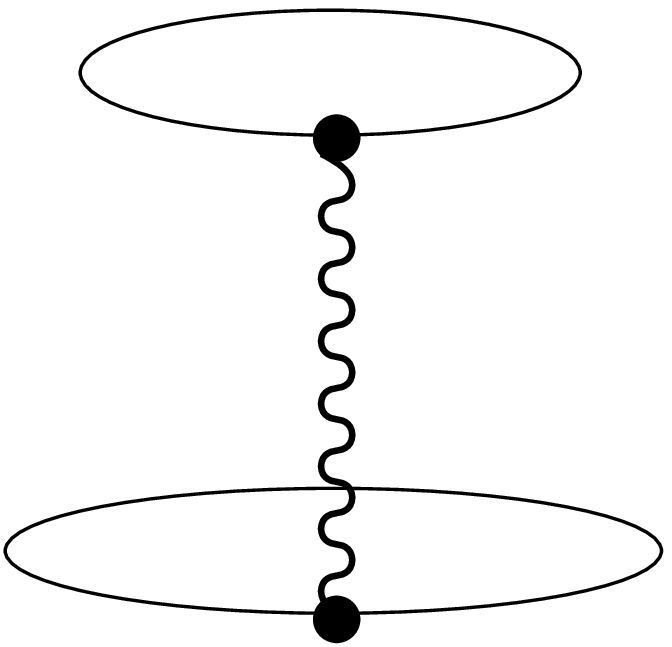} & \cr & & \cr \hline & &\cr
$g^4$: 
&  \feyn{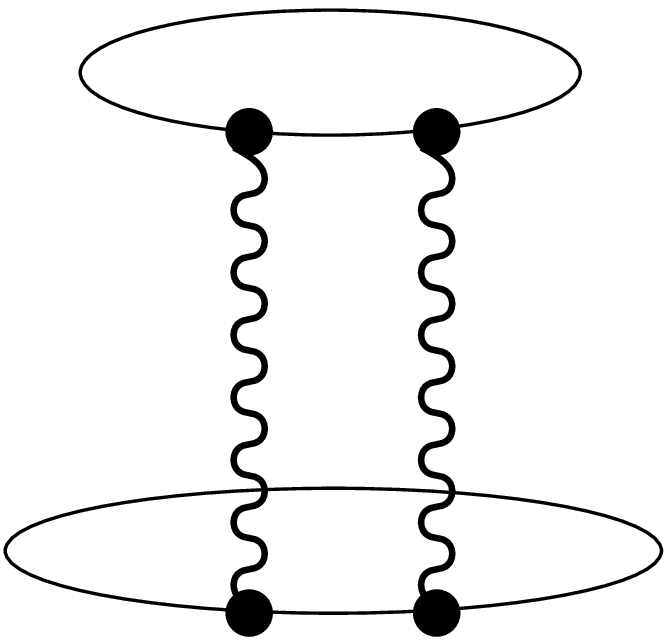} \quad \feyn{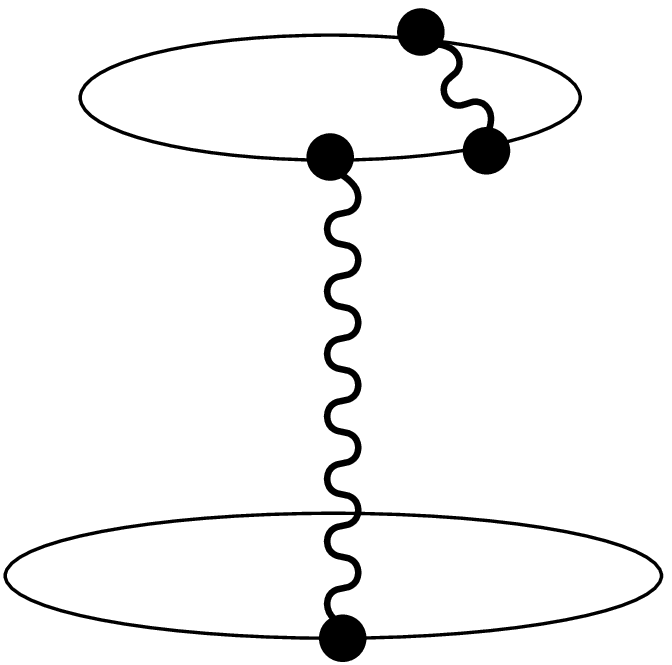} \quad \feyn{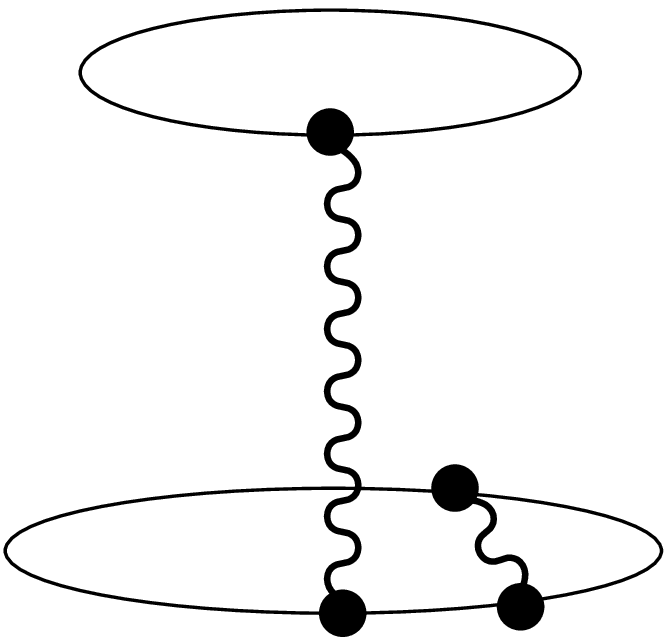} & \cr &&\cr \hline &&\cr
& \feyn{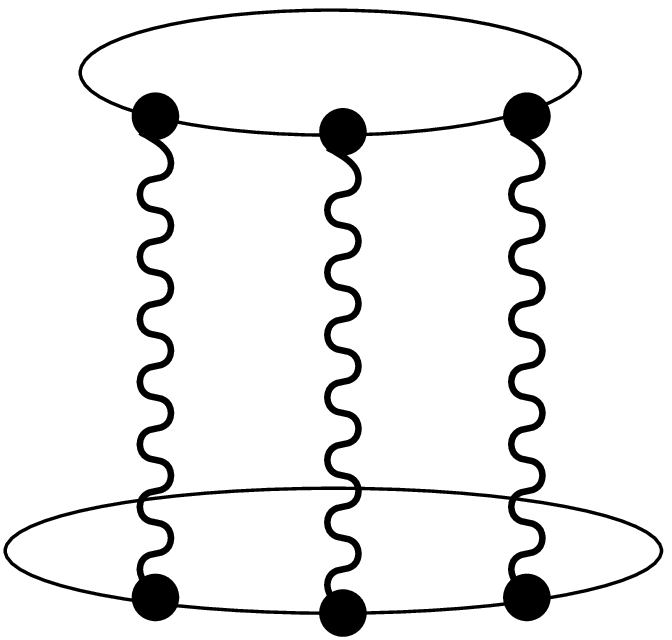} \quad \feyn{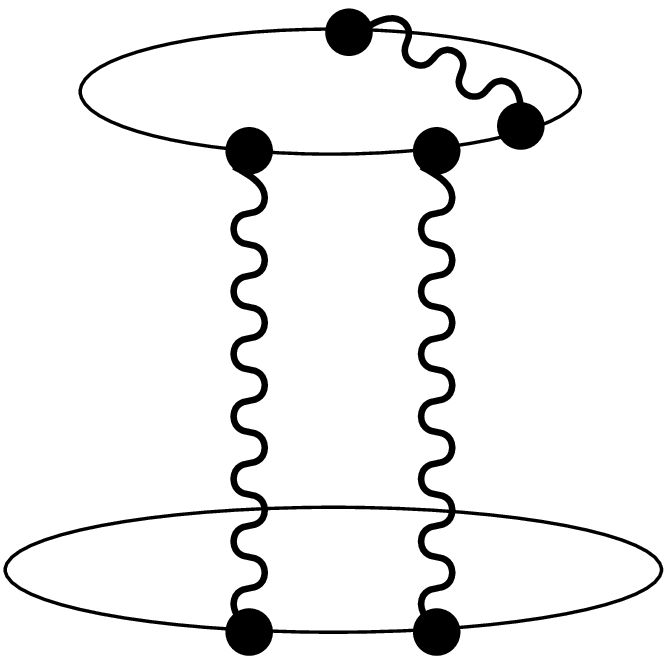}\quad \feyn{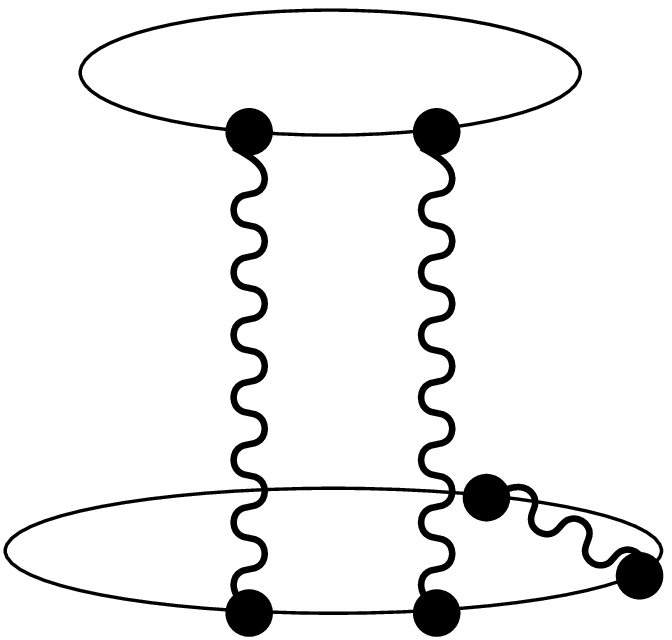}
\quad \feyn{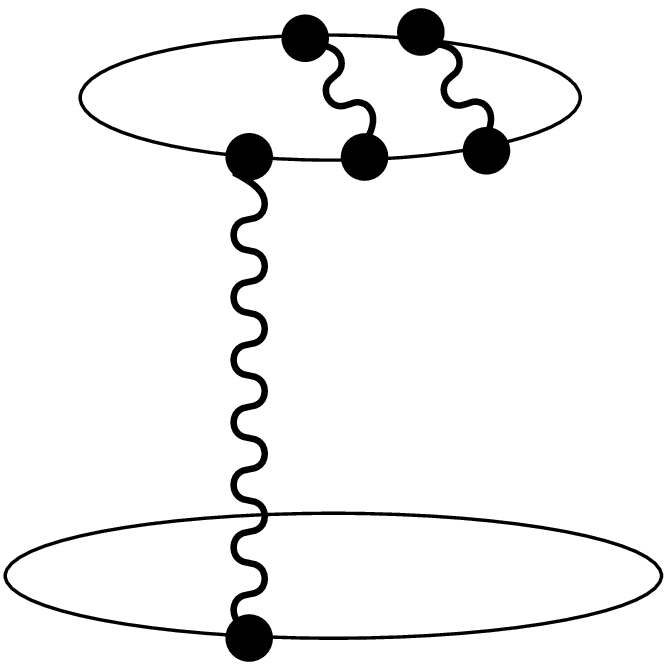} \quad \feyn{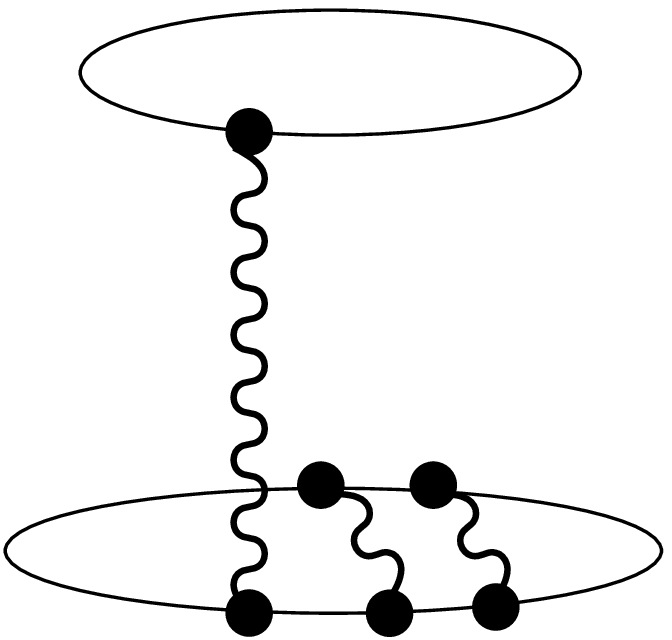}\quad \feyn{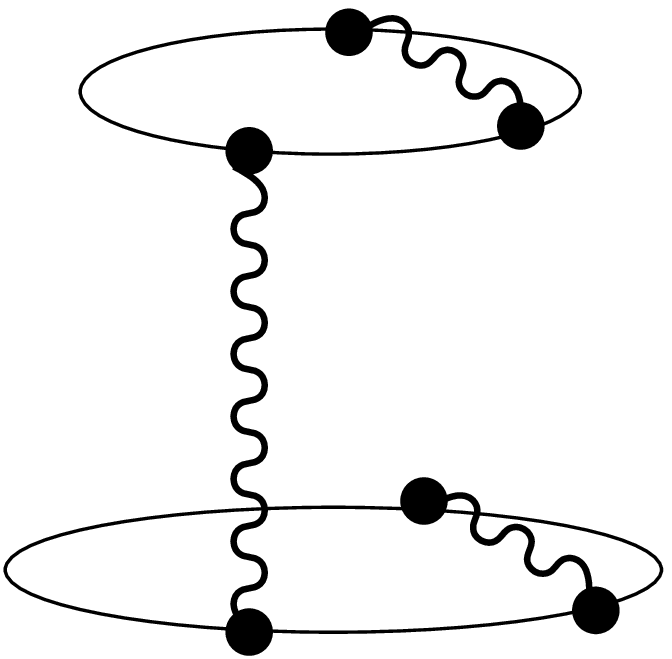} &
\feyn{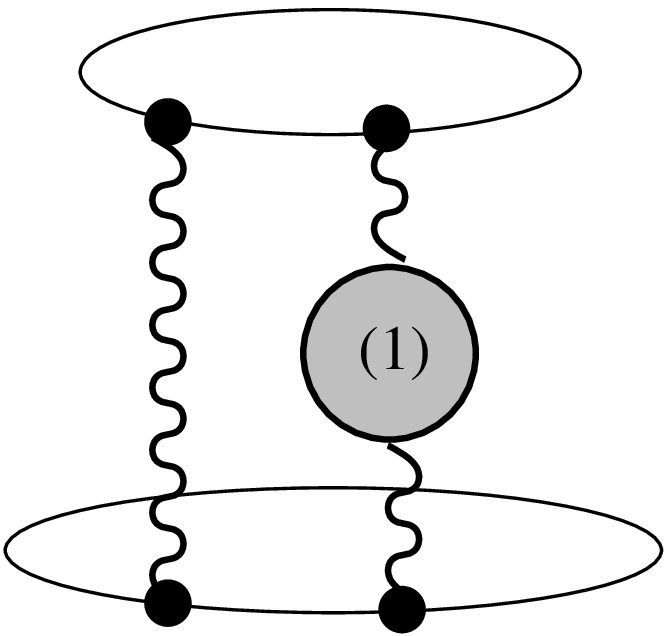} \quad \feyn{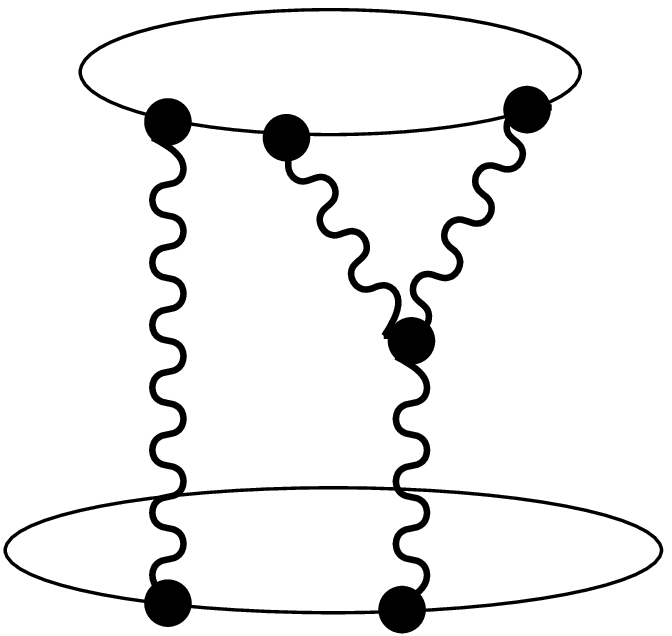} \quad \feyn{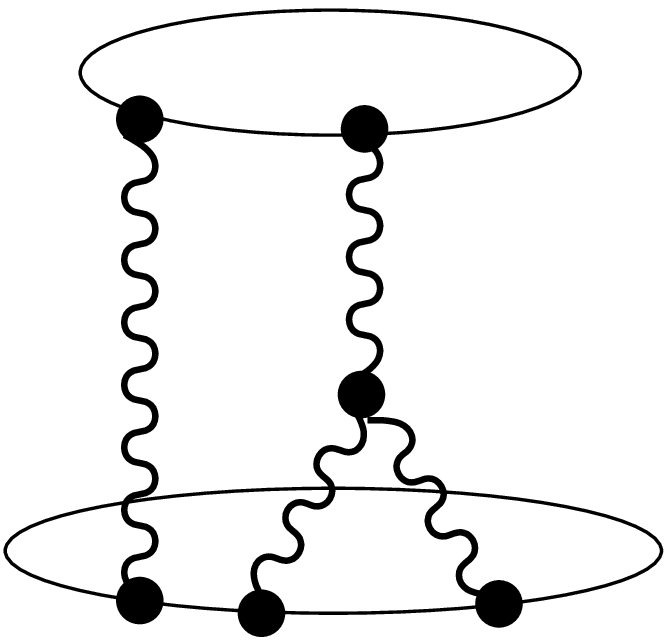} 
\cr  $g^6$: &  &\cr 
& \feyn{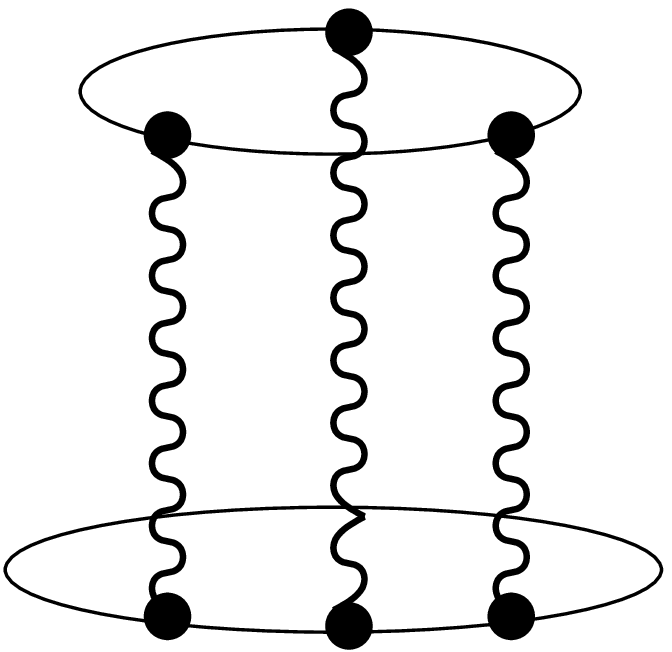} \quad \feyn{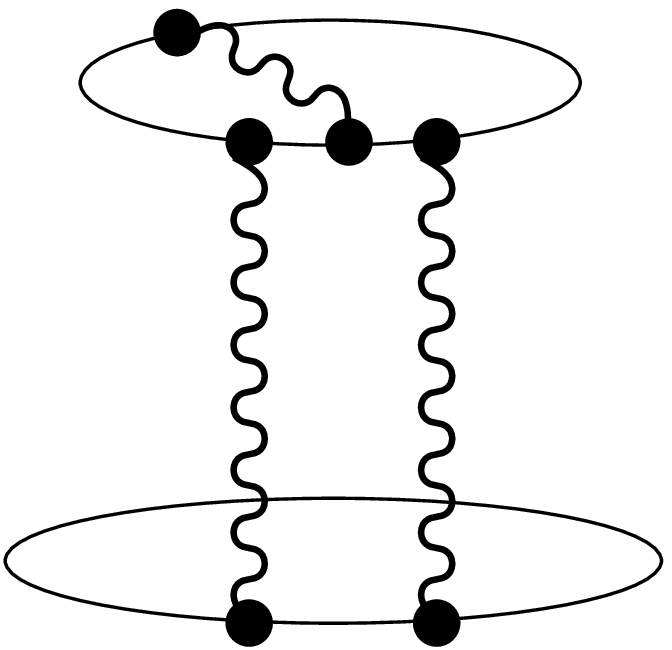} \quad \feyn{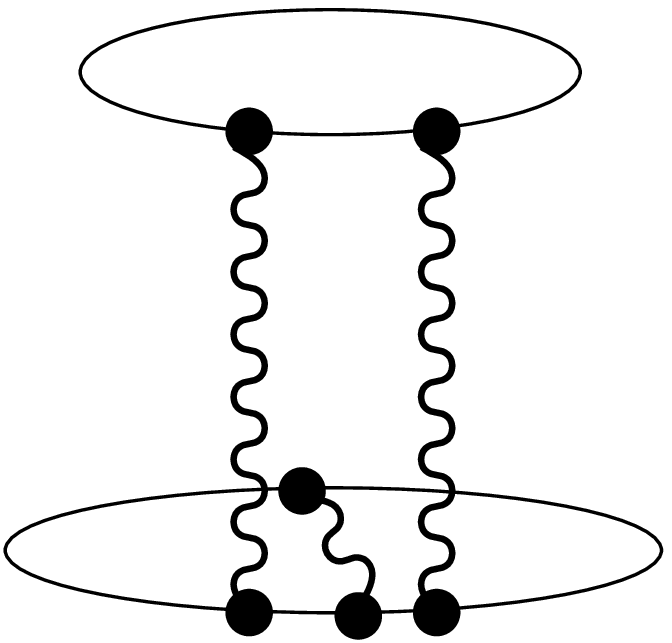} 
\quad \feyn{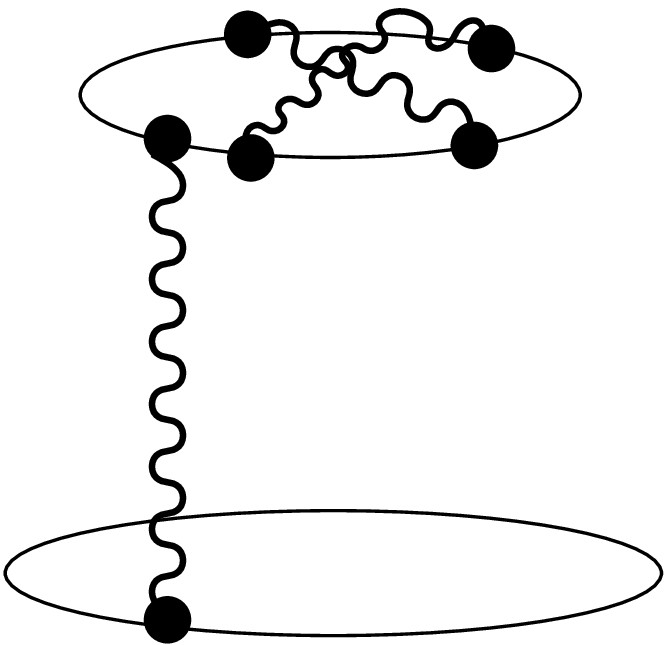}\quad  \feyn{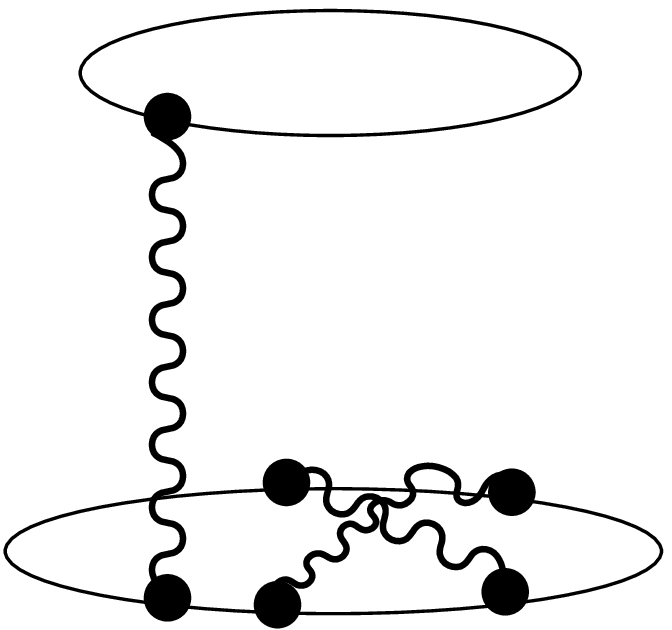}
& \feyn{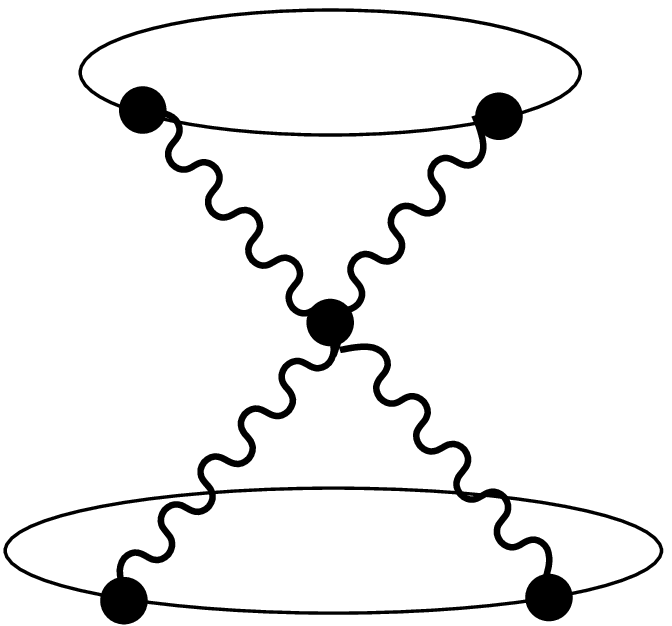}
\quad \feyn{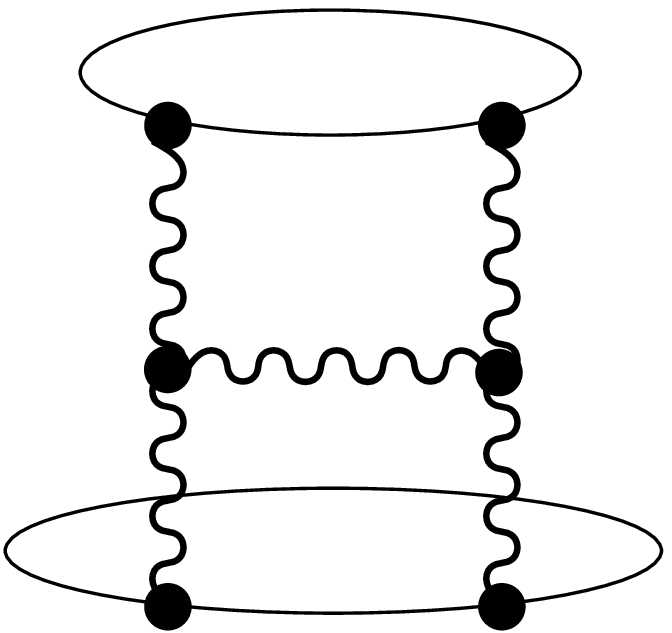} \cr & &\cr
\hline
\end{tabular}
%\end{center}
\caption{The connected  nonvanishing graphs up to order $g^6$. 
The propagators stand for the 
combination of scalar and gluon exchange and
the bubble denotes the
one loop self-energy contribution. We will denote the non-ladder diagrams  
in the above figure by the self-energy, IY, IY${}^T$, X and H-graph
by obvious correspondence. 
Note that the $g^6$ ladder-graphs in the last line
of the figure are non-planar.
}
%\end{figure}
}

We shall be interested in the connected correlator
of {\it two} Maldacena-Wilson loops
\be
\langle \W(C_1)\, \W(C_2) \rangle_{c}
%kleines c is besser um von C_1, C_2 zu unterscheiden ...
= \langle \W(C_1)\, \W(C_2) \rangle-  
\langle \W(C_1)\rangle \, \langle\W(C_2) \rangle
\ee
where we take the curves $C_1$ and $C_2$ to be two parallel, axisymmetric 
circles of opposite 
orientation and, respectively, 
radii $R_1$ and $R_2$ separated by a distance $h$
\bea
x^\mu(\tau)&=& ( R_1\, \cos \tau,\phantom{-} R_1\, \sin\tau,\, 0\, ,\,0)  \nn\\
y^\mu(\sigma)&=& ( R_2\, \cos \sigma, -R_2\, \sin\sigma, \, h \, ,\,0) 
\qquad \tau,\sigma\in [0,2\pi] 
\label{parametrization}
\eea
Performing the diagrammatic expansion in this geometry up to order six in the 
coupling constant $g$ it is seen that a large number of
diagrams vanish identically by taking into account the following two identities
\be
\raisebox{0.25cm}{\feyn{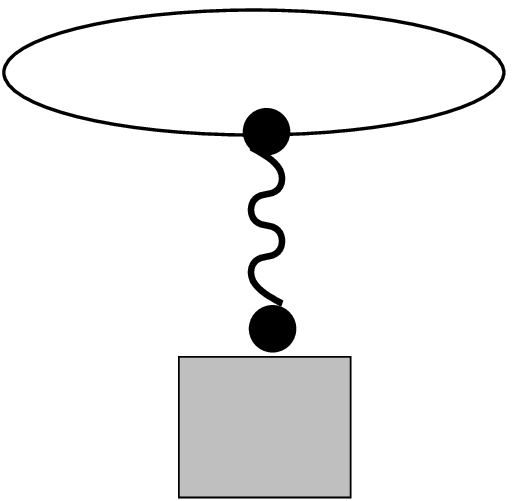}} =0 \qquad 
\feyn{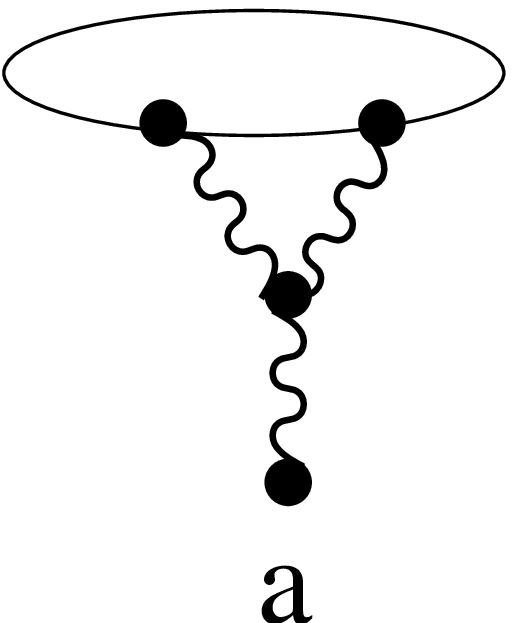} \sim  f_{abc}\, \Tr T^b T^c=0 
\ee
where the box in the first graph stands for any interaction vertex
of the theory and the open leg of the second graph may be contracted 
with either interaction vertices or the second Wilson loop. 
Moreover at order $g^6$ the cancellation of 
\be
\feyn{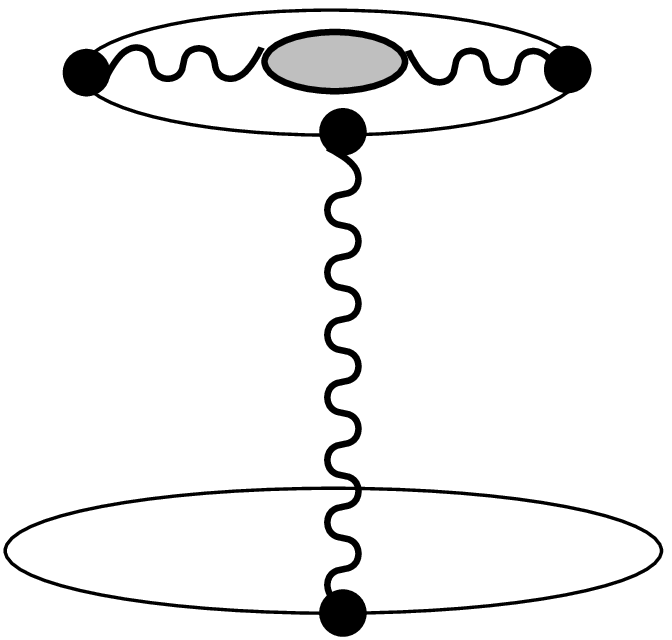} + \feyn{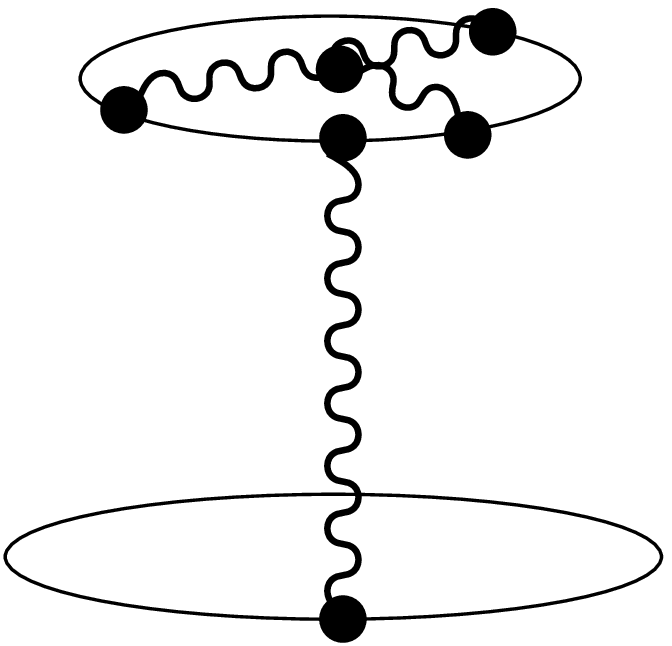}=0
\ee
may be taken into account, a consequence of the cancellation of the
graphs with internal vertices for the circular loop at order $g^4$ shown
in \cite{Erickson:2000af}. The remaining, highly reduced set of
20 non-vanishing planar and non-planar  graphs is depicted in 
the figure.

Let us explicitly compute the leading one ladder graph at order $g^2$. 
The Feynman gauge propagators of the bosonic fields following from \eqn{action}
are
\be
\langle A_\mu^a(x)\, A_\nu^b(y) \rangle =g^2
\frac{ \delta^{ab}\, \delta_{\mu\nu}}{4\pi^2\,(x-y)^2} \qquad
\langle \phi_I^a(x)\, \phi_J^b(y) \rangle = g^2 
\frac{\delta^{ab}\, \delta_{IJ}}{4\pi^2\,(x-y)^2}
\label{prop} \, .
\ee
One then has
\bea
\feyn{1Gluon}&=& -g^2\, \Tr(T^a)\, \Tr(T^a)\, \int_0^{2\pi}\frac{d\tau\, 
d\sigma}{4
\pi^2}\, \frac{\dot x\cdot \dot y - |\dot x|\, |\dot y|}{(x-y)^2}\nn\\
&=& \frac {g^2\, N}{4}\,  \int_0^{2\pi}\frac{d\phi}{2\pi}\,\frac{1+\cos\phi}
{\frac {R_1^2 + R_2^2 + h^2}{2\,R_1\, R_2} -  \cos \phi}
= -\frac{g^2N}{4}\Bigl [ \Wurzel \Bigr ]
\label{1Gluon}
\eea
where we have made use of the parameterization of \eqn{parametrization}.

Turning to the three graphs at $g^4$ one notices that 
there are no path ordering effects at this order, which means that
the computation factorizes into products of one ladder graphs. Taking care
of combinatoric factors one finds
\bea
\feyn{2Gluon}&=& \frac 1 2 \Bigl [ \frac{g^2N}{4}\Bigl ( \Wurzel \Bigr )
\Bigr ] ^2\nn \\
\feyn{2Gluon-a}&=& -\frac{g^4N^2}{32}\, \Bigl[ \Wurzel \Bigr ] \, = 
\feyn{2Gluon-aa}
\label{2Gluon}
\eea
for the diagrams at order $g^4$.

\section{Order $g^6$}

\subsection{The Ladder Graphs}

At order $g^6$ the computation becomes more challenging. Whereas the two and
one ladder diagrams remain easy 
\bea
\feyn{3Gluon-a}&=& \frac{g^6N^3}{3\cdot2^7} \Bigl [ \Wurzel \Bigr ]^2
 \, = \feyn{3Gluon-aa} \nn\\
\feyn{3Gluon-d}&=&  -\frac{g^6N^3}{3\cdot 2^8} \Bigl [ \Wurzel \Bigr ] \, 
= \feyn{3Gluon-dd} \nn\\
\feyn{3Gluon-c}&=& -\frac{g^6N^3}{2^8} \Bigl [ \Wurzel \Bigr ] \nn\\
\feyn{3Gluon-b}&=& \frac{g^6N}{3\cdot 2^8} \Bigl [ \Wurzel \Bigr ]^2
 \, = \feyn{3Gluon-bb} \nn\\
\feyn{3Gluon-nb}&=& -\frac{g^6N}{3\cdot 2^9} \Bigl [ \Wurzel \Bigr ]
\, =\feyn{3Gluon-na}
\label{3Gluon}
\eea
the three ladder graph requires considerably more care. One has
\bea
&\feyn{3Gluon} = {g^6 N^3 \over 3\cdot 2^6} &
\int_0^{2 \pi} 
{d \phi \over 2 \pi}{d \theta_1 \over 2 \pi} {d \theta_2 \over 2 \pi} 
\int_0^{\theta_1} {d \psi_1 \over 2 \pi}
\int_0^{\theta_2} {d \psi_2 \over 2 \pi} \times \nn \\
& &\times {1+\cos \phi \over \kappa -\cos \phi}
~{1+\cos (\psi_1+\theta_2+\phi) \over \kappa -\cos (\psi_1+\theta_2+\phi)}
~{1+\cos (\psi_2+\theta_1+\phi) \over \kappa-\cos (\psi_2+\theta_1+\phi)}
\eea
where we have abbreviated $\kappa := {R_1^2+R_2^2+h^2 \over 2\,R_1 R_2}$.
It is convenient to change to complex coordinates
$(z,z_1,z_2,w_1,w_2)=(e^{i \phi},e^{i \theta_1},e^{i \theta_2},
e^{i \psi_1},e^{i \psi_2})$ and to
introduce the auxiliary parameter $a$ through 
$\kappa :=\frac{1}{2}(a+a^{-1})$. Then the integrations on $w_1,w_2$
are elementary and we obtain a triple contour integral: 
\bea
&\feyn{3Gluon} = {g^6 N^3 \over 3\cdot 2^6 (2 \pi)^2} &
\oint {d z \over 2 \pi i z} {d z_1 \over 2 \pi z_1} {d z_2 \over 2 \pi z_2}
{(z+1)^2 \over (z-a) (z-a^{-1})} \times \nn \\
& &\times
\Big[ \log z_1 - {1+a \over 1-a} \log {z z_1 z_2-a \over z z_2 -a}-
{1+a \over 1-a} \log {1-a z z_2 \over 1-a z z_1 z_2} \Big] \times \nn \\
& &\times
\Big[ \log z_2 - {1+a \over 1-a} \log {z z_1 z_2-a \over z z_1 -a}-
{1+a \over 1-a} \log {1-a z z_1 \over 1-a z z_1 z_2} \Big]
\eea
Carefully keeping track of the cuts introduced by the logarithms
the integrals may be computed term-by-term with the help
of Dilogarithms:
\bea
&\feyn{3Gluon} = \frac{g^6 N^3}{12} & 
\Bigg\{ \,\Bigl[-\frac{1}{4} \Bigl(\Wurzel\Bigr)\, \Bigr]^3\nn\\&&\quad
-\frac{3}{2^5 \pi^2} \Bigl[\Wurzel \Bigr]\,
\ft{(R_1+R_2)^2+h^2}{(R_1-R_2)^2 + h^2}\,
{\rm Li}_2(a^2) \Bigg\}\label{ladderfinal}\\
& {\rm where} & a^2=
\frac{\left(h^2+R_1^2+R_2^2-\sqrt{(h^2+R_1^2+R_2^2)^2-4 R_1^2 R_2^2} 
\right )^2}{ 4 R_1^2 R_2^2} 
\nn
\eea 
The non-planar cousin of the three ladder diagram is obtained by
the same methods and reads
\bea
\feyn{3Gluon-NP}&=& \ft{g^6 N}{12} 
\Bigg\{ \, \Bigl[-\frac{1}{4} \Bigl(\Wurzel\Bigr) \Bigr]^3 \nn\\&&\qquad
+\frac{3}{2^5 \pi^2} \Bigl[\Wurzel \Bigr]\,
\ft{(R_1+R_2)^2+h^2}{(R_1-R_2)^2 + h^2}\,
{\rm Li}_2(a^2) \Bigg\} \, .
\label{ladderfinalnp}
\eea

\subsection{The Self-Energy Graph}

${\cal N}=4$ gauge theory is a finite theory. This, however, does not
mean that gauge dependent, individual Feynman diagrams are finite 
as well. An important example are the self energies of the gauge bosons
and scalars, which are infinite in four dimensions.
In order to isolate the divergences we use regularization
by dimensional reduction which maintains supersymmetry. This procedure 
considers supersymmetric Yang-Mills theory in $2\omega$ dimensions 
as a dimensional reduction of the ten dimensional model. One hence has
a $2\omega$ component gauge field $A_\mu^a$, $10-2\omega$ scalars
and a 16 component fermion field at every stage of the computation. 
The one-loop self energy of the vector and scalar fields was computed
in this regularization in \cite{Erickson:2000af}. In configuration
space one has
\bea
\raisebox{-0.2cm}{\epsfxsize=2cm\epsfbox{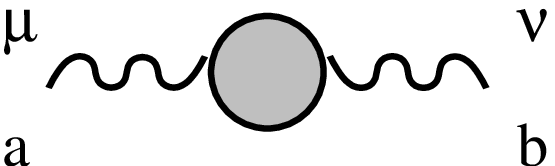}}&=& \delta_{\mu\nu} 
\, \delta^{\tilde a \tilde b}\, g^4 N\, A_\omega[x]
+ \frac{\delta^{\tilde a \tilde b}\, g^4N\,\Gamma^2(\omega-1)}
{2^7\pi^{2\omega}\, (\omega-2)^2\, (\omega-3)\,(2\omega-3)}\,
\partial_{x^\mu}\partial_{x^\nu} \Bigl (
[x^2]^{4-2\omega} \Bigr )\nn\\
\raisebox{-0.2cm}{\epsfxsize=2cm\epsfbox{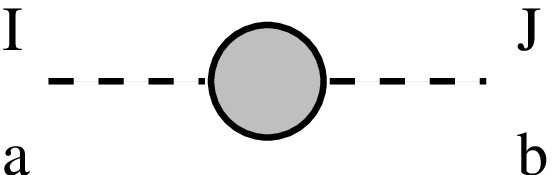}}&=&
\delta_{IJ}\, \delta^{\tilde a \tilde b}\, g^4 N\, A_\omega[x]
\label{1loopprop}
\eea
with group indices $\tilde a,\tilde b=1,\ldots, N^2-1$ in the 
interacting $SU(N)$ sector and where
the universal function $A_\omega[x]$ is given by 
\be
A_\omega[x]=
\frac{\Gamma^2(\omega-1)}{2^5\pi^{2\omega}\, (2-\omega)\,
(2\omega-3)}\, \frac{1}{[x^2]^{2\omega-3}} 
\ee
which diverges in four dimensions.

Turning toward the evaluation of the self-energy graph itself we find
\bea
\feyn{SE}&=& \frac{4\, g^6 N}{2!\,2!}\,\Bigl [\Tr(T^{\tilde a}\, T^{\tilde b})
\Bigr ]^2\,
\int_0^{2\pi} d\tau_1\, d\sigma_1\, (\xd{1}\cdot\yd{1} -|\xd{1}|\, |\yd{1}|)
\Delta_\omega[x_1(\tau_1)-y_1(\sigma_1)] \nn\\
&& \qquad\times\int_0^{2\pi} d\tau_2\, d\sigma_2\,
(\xd{2}\cdot\yd{2} -|\xd{2}|\, |\yd{2}|)\, 
A_\omega[x_2(\tau_2)-y_2(\sigma_2)]
\label{SE}
\eea
Here $\Delta_\omega$ denotes the propagator \eqn{prop} in $2\omega$
dimensions with stripped off coupling constant and Kronecker delta-functions
\be
\Delta_\omega(x)=\frac{\Gamma(\omega-1)}{4\pi^\omega}\, 
\frac{1}{[x^2]^{\omega-1}}\, .
\label{dimregprop}
\ee
We stress that the double derivative term in the vector self-energy
of \eqn{1loopprop} drops out as it constitutes a
total derivative here. However, at higher orders this inhomogenous 
contribution to the one loop self-energy might become relevant. 
The self-energy graph \eqn{SE} is divergent in 4d --
this divergence will cancel against the IY graphs. We leave it as 
it stands for the time being, noting that the prefactor in \eqn{SE}
takes the value $g^6N(N^2-1)/4$.

\subsection{The IY-Graphs: Cancellation of Divergences}

When evaluating the two IY-diagrams we encounter a second type
of infinity distinct in origin from the type of ``bulk'' divergences 
discussed in the last subsection. When two of the three legs of the
three-vertex come close on the boundary, the confluence causes a logarithmic
``boundary'' divergence. To be consistent, we regulate it by the same
procedure of dimensional reduction used above. We then find that
it cancels against the self-energy divergence. 
One furthermore needs to carefully extract the finite contribution next to 
the infinite piece, which is subtle and technically quite involved. 
In fact, we will only do this explicitly for the special case $R_1=R_2$
-- the finite part for the case of more general geometries is not
needed for reaching our physical conclusions.
Using the parameterization of \eqn{parametrization}
we denote the three points
on the upper loop by $x_i(\tau_i)$ ($i=1,2,3$) and the two 
points on the lower loop by $y_j(\sigma_j)$ ($j=1,2$). 
Path ordering only occurs for the upper loop in the $\tau_i$ parameters.
Performing the usual Wick contractions one is confronted with
the integral
\bea
\feyn{IY} &=& \ft{g^6\, N\, (N^2-1)}{8} \int_0^{2\pi}\!\!
d\tau_1 d\tau_2 d\tau_3\,
\int_0^{2\pi}\!\!d\sigma_1d\sigma_2\, \epsilon(\tau_1,\tau_2,\tau_3)\,
(\xd{3}\cdot\yd{1}-R_1\,R_2)\, \Delta_\omega(x_3-y_1) \nn\\
&&\times \Bigl\{ (\xd{1}\cdot \yd{2}-R_1\, R_2)\, [\xd{2}\cdot
\partial_{y_2}-\xd{2}\cdot\partial_{x_1}] %\nn\\&&\quad 
 - (\xd{1}\cdot \xd{2}-R_1^2)\, \yd{2}\cdot
\partial_{x_2}\, \Bigr \}\, G(x_1,x_2,y_2) \nn\\&&
\label{IY}
\eea
where an ordering symbol has been introduced obeying 
$\epsilon(\tau_1,\tau_2,\tau_3)=1$ for $\tau_1>\tau_2>\tau_3$
and antisymmetric under any permutation of $\tau_i$. 
Moreover we have defined the dimensional regulated 
three-point function
\be
G(x_1,x_2,y_2)=\int d^{2\omega}r\, \Delta_\omega(x_1-r)\, 
\Delta_\omega(x_2-r)\, \Delta_\omega(y_2-r) 
\label{G3}
\ee
with $\Delta_\omega(x)$ given by \eqn{dimregprop}.
As the ``I''-leg term $(\xd{3}\cdot\yd{1}-R_1\,R_2)\, \Delta_\omega(x_3-y_1)$
depends homogeneously on the angular combination
$(\sigma_1+\tau_3)$ a shift in the 
lower $\sigma_1$ integration by $\sigma_1\rightarrow \sigma_1-\tau_3$
lets the ``I'' leg of the IY-graph decouple.
Then the only dependence on $\tau_3$ in the integrand of
\eqn{IY} sits in the ordering symbol $\epsilon(\tau_1,\tau_2,\tau_3)$
and this integral may be performed to yield
\be
\int_0^{2\pi}d\tau_3 \, \epsilon(\tau_1,\tau_2,\tau_3) = 2\pi\,
\sign(\tau_1-\tau_2)-2(\tau_1-\tau_2) \equiv E(\tau_1.\tau_2)\, .
\label{E}
\ee
In order to proceed one now introduces Feynman parameters for
\eqn{G3} and performs the integral over $r$. 
Collecting the dependence on the ``I''-leg and the
prefactor of \eqn{IY} in the symbol (I),
\be
\mbox{(I)}\equiv \frac{g^6\, N\, (N^2-1)}{16\pi} \, 
\int_0^{2\pi} d\tau\, d\sigma\, (\dot x\cdot\dot y - R_1R_2)
\Delta_\omega[x(\tau)-y(\sigma)]
\label{constant}
\ee
the expression \eqn{IY} takes the rather complicated form
\bea
\feyn{IY}&=&
\mbox{(I)}\,\frac{\Gamma(2\omega-2)}{2^5\pi^{2\omega}}
\int_0^{2\pi}\!\!d\tau_1 d\tau_2d\sigma_2 \int_0^1d\alpha d\beta d\gamma
\, (\alpha\beta\gamma)^{\omega-2}\, \delta(\alpha+\beta+\gamma-1)
\, E(\tau_1,\tau_2)\nn\\
&& \!\!\!\!\!\!\!\!\!\!\!\!\!\!\!
\times\Bigl\{- R_1R_2\left(1+\cos(\sigma_2+\tau_1)\right)
\, \Bigl [ R_1R_2\sin(\sigma_2+\tau_2)\, (2\alpha+\beta)\gamma
+ R_1{}^2\sin\tau_{12}\, (2\gamma+\beta)\alpha\Bigr ]\nn\\
&&
 + R_1{}^3R_2 (1-\cos\tau_{12})\, \sin(\sigma_2+\tau_2)\,
(2\alpha+\gamma)\beta\Bigr\}\,\frac{1}{\Delta^{2\omega-2}}\,.
\label{IY1}
\eea
where $\tau_{12}=\tau_1-\tau_2$ and
\bea
\Delta&=&(1-\gamma)\gamma (R_1{}^2+R_1{}^2+h^2)+ 2\alpha\beta\,
R_2{}^2(1-\cos\tau_{12})\nn\\&&\qquad
-2R_1R_2\Bigr(\alpha\gamma\cos(\sigma_2+\tau_1)
+\beta\gamma\cos (\sigma_2+\tau_2)\Bigl)
\eea
In order to split off the divergent part of the integral we make use of
the following key identity for the integrand of \eqn{IY1} 
\bea
&&
E(\tau_1,\tau_2)\Bigl\{- R_1R_2\left(1+\cos(\sigma_2+\tau_1)\right)
\, \Bigl [ R_1R_2\sin(\sigma_2+\tau_2)\, (2\alpha+\beta)\gamma\nn\\&&
+ R_1{}^2\sin\tau_{12}\, (2\gamma+\beta)\alpha\Bigr ]
 + R_1{}^3R_2 (1-\cos\tau_{12})\, \sin(\sigma_2+\tau_2)\,
(2\alpha+\gamma)\beta\Bigr\}\,\frac{1}{\Delta^{2\omega-2}}=\nn\\&&
 -\ft 1 2 R_1{}^2\,\partial_{\tau_2}\left ( E(\tau_1,\tau_2)\,
\frac{1-\cos\tau_{12}}{\Delta^{2\omega-3}}\right ) + R_1R_2\,
\partial_{\tau_1}\left (  E(\tau_1,\tau_2)\,
\frac{1+\cos(\sigma_2+\tau_1)}{\Delta^{2\omega-3}}\right )\nn\\&&
+\ft 12  R_1R_2\,
\partial_{\tau_2}\left ( E(\tau_1,\tau_2)\,
\frac{1+\cos(\sigma_2+\tau_1)}{\Delta^{2\omega-3}}\right )
+(\mbox{IY})_{\mbox{\tiny SE}}+(\mbox{IY})_{1}+(\mbox{IY})_{2}
+(\mbox{IY})_{\omega-2}\nn\\&&
\label{IY2}
\eea
where
\bea
&(\mbox{IY})_{\mbox{\tiny SE}}=& \frac{2\pi\delta(\tau_{12})}{2\omega-3}\, 
R_1R_2\,\frac{1+\cos(\sigma_2+\tau_1)}{\Delta^{2\omega-3}} \nn\\
&(\mbox{IY})_{1}=&
-\frac{1}{2\omega-3}\, 
\left\{R_1{}^2\,\frac{1-\cos\tau_{12}}{\Delta^{2\omega-3}}+
R_1R_2\,\frac{1+\cos(\sigma_2+\tau_1)}{\Delta^{2\omega-3}} \right \}\nn\\
&(\mbox{IY})_{2}=&
\frac{E(\tau_1,\tau_2)}{\Delta^{2\omega-2}}\, \left [ (R_1+R_2)^2+h^2\right ]
\, \gamma\, (1-\gamma)\, \left [ R_1R_2\, \sin(\sigma_2+\tau_1)-
\ft 1 2 R_1{}^2\, \sin\tau_{12}\right ] \nn\\
&(\mbox{IY})_{\omega-2}=&
\frac{2\omega-4}{2\omega-3}\, \frac{E(\tau_1,\tau_2)}{\Delta^{2\omega-3}}
\, \left \{ \ft 1 2 R_1{}^2\, \sin\tau_{12}-R_1R_2\, \sin(\sigma_2+\tau_1)
\right \} 
\eea
which holds under integration over the angles $\tau_1,\tau_2,\sigma_2$ and the
Feynman parameters $\alpha,\beta,\gamma$. Plugging this relation back into
\eqn{IY1} one sees that the total derivative terms of the right-hand-side 
of the identity \eqn{IY2} drop out. 
Moreover one can show that the integral over $(\mbox{IY})_{\omega-2}$ is finite in
$2\omega=4$ dimensions, therefore  due to the $(2\omega-4)$ prefactor 
this term drops out as well.

To continue let us investigate the contribution of
the $(\mbox{IY})_{\mbox{\tiny SE}}$ term coupling to the
$\delta$-function.
Here the Feynman parameter integral factorizes and may be performed
to yield
\bea
\int (\mbox{IY})_{\mbox{\tiny SE}} &=&
2
\pi\,\mbox{(I)}\int_0^{2\pi}d\sigma_2 d\tau_1 \frac{\Gamma(2\omega-2)\, R_1R_2}
{2^5\pi^{2\omega}(2\omega-3)}
\, \frac{1+\cos(\sigma_2+\tau_1)}
{[R_1{}^2+R_2{}^2+h^2-2R_1R_2\,\cos(\sigma_2+\tau_2)]^{2\omega-3}}
\nn\\
&& \quad \times
\int_0^1d\alpha d\beta d \gamma\,
\frac{ (\alpha\beta\gamma)^{\omega-2}\, \delta(\alpha+\beta+\gamma-1)}
{(\gamma(1-\gamma))^{2\omega-3}}\nn\\
&=&
-2\pi \,\mbox{(I)}\,
 \frac{\Gamma^2(\omega-1)}{2^5\pi^{2\omega}\, (2-\omega)\, (2\omega-3)}
\int d\sigma_2 d\tau_1\, \frac{\xd{1}\cdot\yd{2}-R_1R_2}
{[(x_1-y_2)^2]^{2\omega-3}}= -\frac{1}{2}\, \feyn{SE}
\eea
where we have undone the parameterization of \eqn{parametrization} in the
second step. In the last step we see that upon using \eqn{constant}
this part of the IY-graph precisely
cancels half of the self-energy graph \eqn{SE}. The second half comes from the
mirror IY${}^T$-graph which may be obtained from the above by swapping
$R_1\leftrightarrow R_2$. To summarize we thus have
\be
\feyn{IY}=-\ft 1 2\, \feyn{SE} + \int (\mbox{IY})_{1} + \int (\mbox{IY})_{2}
\ee 
in a symbolic notation.

For the evaluation of the $(\mbox{IY})_{1}$ contribution
we go back to exactly $2 \omega=4$ dimensions:
$(\mbox{IY})_{1}$ turns out to be finite.
We found it useful to reintroduce the integral over $r$ in \eqn{G3} again
and remove the Feynman parameters. One may then treat the angular integrals
by contour techniques and is left with an integral over the space-point
$r$. Doing this one starts with
\be
\int (\mbox{IY})_{1} = -\frac{\mbox{(I)}}{2^5\pi^6}\, 
\int_0^{2\pi}\!\!d\tau_1d\tau_2d\sigma_2
\, \int d^4r \, \frac{\xd{1}\cdot\xd{2}-R_1{}^2+\xd{1}\cdot
\yd{2}-R_1R_2}{(x_1-r)^2\,(x_2-r)^2\,(y_2-r)^2}\, .
\label{IY11}
\ee
Noting that in four dimensions the symbol $\mbox{(I)}$ takes the value
\be
\mbox{(I)}_{\omega=2}=\frac{g^6N(N^2-1)}{32\pi}\,\left [\, \Wurzel\,\right ]
\ee
and performing the angular integrals leads to the final result
\bea
\int (\mbox{IY})_{1} &=&
-\frac{g^6N(N^2-1)}{256\pi^3}\,\left [\, \Wurzel\,\right ]
\, \int_0^\infty d\rho \int_{-\infty}^\infty dr_3\int_{-\infty}^\infty dr_4
\frac{1}{ \rho\, L_1\, \sqrt{L_2^h}}\nn\\
&& \times\Bigl [\rho^4 + h^2\,(\sqrt{L_1} - \rho^2 - R_1^2 - r_3^2 - r_4^2) 
+  2\, h\,r_3\,(-\sqrt{L_1} + \rho^2 + R_1^2 + r_3^2 + r_4^2) \nn\\
&& \quad
- 
  (\sqrt{L_1} - R_1^2 - r_3^2 - r_4^2)\,(\sqrt{L_2^h} + 2\,R_1^2 
- R_2^2 + r_3^2 + r_4^2) \nn\\
&& \quad - 
  \rho^2\,(\sqrt{L_1} - \sqrt{L_2^h} + 5\,R_1^2 + 4\,R_1\,R_2 + R_2^2 
- 2\,(r_3^2 + r_4^2))
\Bigr ] 
\label{IY1Final}
\eea
where we have introduced
\bea
L_1&=& \left[(\rho-R_1)^2+r_3^2+r_4^2\right]\,
\left[(\rho+R_1)^2+r_3^2+r_4^2\right] \nn\\
L_2^h &=& \left[(\rho-R_2)^2+(r_3-h)^2+r_4^2\right]\,
\left[(\rho+R_2)^2+(r_3-h)^2+r_4^2\right] \, .
\label{Pdef}
\eea
Note that due to the symmetry of our loop configuration in the $1-2$ plane one
is able to reduce \eqn{IY11} to a three dimensional integral. This is as
far as one can get analytically -- \eqn{IY1Final} may be evaluated with
good precision numerically. We will return to an analytical treatment of
 \eqn{IY1Final} in the potential limit.

Turning to the $(\mbox{IY})_{2}$ contribution one finds the following
compact expression via a derivative with respect to $h^2$
\be
\int (\mbox{IY})_{2} = \frac{ \mbox{(I)}}{2^5\pi^6}\, 
\frac{\partial}{\partial h^2}\, \left [
\int_0^{2\pi}\!\!d\tau_1d\tau_2d\sigma_2\, E(\tau_1,\tau_2)
\, \int d^4r \, \frac{\yd{2}\cdot x_{1}+\ft 1 2 \xd{2}\cdot
x_1}{(x_1-r)^2\,(x_2-r)^2\,(y_2-r)^2}\, \right ]\,
\label{IY12}
\ee
after reintroducing the $r$ integration. Now the integration
over the angles is less straightforward due to the path ordering
symbol $E(\tau_1,\tau_2)$. The calculation proceeds along lines similar
to the three-ladder case: One introduces complex variables
$(z,w_1,w_2)=(e^{i \sigma_2},e^{i \tau_1},e^{i \tau_2})$
and rewrites the loop part of eq.(\ref{IY12}) 
as a multiple, open contour integral:
\bea
& \int (\mbox{IY})_{2} = &\frac{ \mbox{(I)}}{2^6\pi^6}\,{1 \over R_1 R_2}
\int d^4r 
\,{1 \over r_1-i r_2}\, {1 \over r_1^2+r_2^2} \,
\frac{\partial}{\partial h^2}\, 
\oint {d z \over i z} \oint {d w_2 \over i w_2}
\left \{ \quad \right \} \nn \\
& {\rm where} & \nn \\
&\left \{ \quad \right \}  = &
 \left [
- 4 \pi i \int_1^{w_2} {d w_1 \over i w_1} +
\oint {d w_1 \over i w_1} \, (\log w_1 -\log w_2) \right ] \times \nn \\
& & \times 
{R_1 \, z\, (w_2^2-w_1^2) + 2 \, R_2 \, w_2 \, (w_1^2 z^2-1) \over
(z-z_-)(z-z_+)(w_1-w_-)(w_1-w_+)(w_2-w_-)(w_2-w_+) } 
\eea
Here $z_\pm$,$w_\pm$ are, respectively,
the roots of the quadratic equations
\bea
z^2-\frac{1}{R_2} \, {R_2^2+ \rho^2 +(r_3-h)^2+r_4^2 \over r_1+i r_2}\; z+
{r_1-i r_2 \over r_1 + i r_2} =0 \nn \\
w^2-\frac{1}{R_1} \, {R_1^2+ \rho^2 +r_3^2+r_4^2 \over r_1-i r_2} \; w+
{r_1+i r_2 \over r_1 - i r_2} =0
\eea
Now the integrations on $w_2$,$w_1$ are elementary, if tedious, and the
$z$-integration can be performed by the residue theorem. The final
expression is
\bea
\int (\mbox{IY})_{2}&=&
\frac{g^6N(N^2-1)}{64\pi^3}\,\left[\,\Wurzel\,\right ]\,
\left[h^2 + (R_1 + R_2)^2)\right]\nn\\
&&\times
 \int_0^\infty d\rho \int_{-\infty}^\infty dr_3\int_{-\infty}^\infty dr_4
\,\frac{h-r_3}{h\,\rho\,\sqrt{L_1}(L_2^h)^{3/2}}\,
\Bigl\{\rho^4 + h^2\,(R_1^2 + \rho^2 + r_3^2 + r_4^2) \nn\\
&&\quad- 
2\, h\,r_3\,(R_1^2 +\rho^2 + r_3^2 + r_4^2) + (R_1^2 + r_3^2 + r_4^2)\,
(R_2^2 + r_3^2 + r_4^2)\nn\\
&& \quad + \rho^2\,\Bigl(R_1^2 - 3\,R_2^2 + 2\,(r_3^2 + r_4^2)\Bigr )\Bigr \}\,
\nn\\&&\quad\times\,
\log\, \left[2 + \frac{\sqrt{L_1}\,(R_1^2 +\rho^2 + r_3^2 + r_4^2)
 - ( R_1^2+\rho^2  + r_3^2 + r_4^2)^2}{{2\rho^2\,R_1^2}}\right]
\label{IY2Final}
\eea
which completes the computation of the IY-graphs. 

As already mentioned above, this integral still has a residual
logarithmic divergence unless
$R_1=R_2$. However, one can show that for $R_1 \neq R_2$ this
divergence is precisely canceled by the 
analogous divergence coming from $(\mbox{IY}^T)_{2}$.

\subsection{The X-Graph}

The next graph on our list is the X-graph. Let us denote the two points on the 
upper loop by $x_i(\tau_i)$ and the two points on the lower loop
by $y_i(\sigma_i)$ ($i=1,2$). Happily no path ordering is required and the 
graph turns out to be finite in four dimensions. After taking care
of combinatorial factors and performing the usual Wick contractions
one arrives at
\bea
\feyn{X}&=& \ft{g^6N(N^2-1)}{8}\, \int_0^{2\pi}d\tau_1\, d\tau_2\,
d\sigma_1\, d\sigma_2\nn\\&&
 \times \Bigl [ (\xd{1}\cdot\yd{2}-R_1\, R_2)\, 
(\yd{1}\cdot\xd{2}-R_1\, R_2)
 -  (\xd{1}\cdot\xd{2}-R_1{}^2)\,(\yd{1}\cdot\yd{2}-R_2{}^2)\,
\Bigr ]\nn\\&&\quad \times
 \ft{1}{(4\pi^2)^4}\, \int d^4 r \frac{1}{(x_1-r)^2\, (x_2-r)^2\,
(y_1-r)^2\,(y_2-r)^2}
\eea
Leaving the $r$ integral untouched for the moment, we first perform
the four angular integrals over $\tau_i$ and $\sigma_i$ which are all
elementary. One then obtains the following compact result
\bea
\feyn{X}&=& \frac{g^6N(N^2-1)}{256\pi^3}\, 
 \int_0^\infty d\rho \int_{-\infty}^\infty dr_3\int_{-\infty}^\infty dr_4
\,\frac{1}{\rho\,L_1\,L_2^h}\, \Bigl[ -h^2\,R_1 + 
\sqrt{L_2^h}\,R_1 
\nn\\
&&\quad
 + \sqrt{L_1}\,R_2+ 2\,h\,R_1\,r_3 - 
 (R_1 + R_2)\,(\rho^2 + R_1\,R_2 + r_3^2 + r_4^2)\,\Bigr]^2
\label{XFinal}
\eea
where we again made use of the abbreviations \eqn{Pdef} for 
$L_1$ and $L_2^h$. The integrations over $\rho$, $r_3$ and $r_4$ 
appear to be not expressable in terms of known functions. 

\subsection{The H-Graph}

For the computation of the H-graph we denote the two points on the 
upper loop by $x_i(\tau_i)$ and the two points on the lower loop
by $y_i(\sigma_i)$ ($i=1,2$). Again there is no path ordering for this
graph, and it is completely finite due to cancellations between gauge
and scalar degrees of freedom, allowing us to work in four dimensions
from the outset. One then starts off with the following integral
\bea
\feyn{H}&=&\ft{g^6N(N^2-1)}{8}\, \int_0^{2\pi}d\tau_1\, d\tau_2\,
d\sigma_1\, d\sigma_2\nn\\
&&\times \left [ 2\,\yd{1}^M\, \xd{1}\cdot\partial_{y_1}
-2\,\xd{1}^M\, \yd{1}\cdot\partial_{x_1} + (\xd{1}\cdot\yd{1}-R_1R_2)\, 
(\partial_{x_1^M}
-\partial_{y_1^M})\, \right ] \nn\\
&&\times
\left [ 2\,\yd{2}^M\, \xd{2}\cdot\partial_{y_2}
-2\,\xd{2}^M\, \yd{2}\cdot\partial_{x_2} + (\xd{2}\cdot\yd{2}- R_1R_2)\, 
(\partial_{x_2^M}
-\partial_{y_2^M})\, \right ]\\ &&
\times
\ft{1}{(4\pi^2)^5}\, \int d^4z\, d^4w\, \frac{1}{(x_1-z)^2\,(y_1-z)^2\,
(z-w)^2\, (x_2-w)^2\,(y_2-w)^2} \nn
\eea
where we have made use of a handy five dimensional index $(M=\mu,5)$
notation where $\xd{i}^M=(\xd{i}^\mu,|\xd{i}|)$ and $\partial_{x_i^M}=
(\partial_{x_i^\mu},0)$. As usual we will now perform the angular
integrals $(\tau_1,\tau_2,\sigma_1,\sigma_2)$ and leave the space-integrals
over $z$ and $w$ untouched. The angular integration 
factorizes and one is left with the computation of the five dimensional vector 
$H^M(z)$ 
\bea
H^M(z)&\equiv& \int_0^{2\pi}d\tau\, d\sigma\,
\left [\,2\,\yd{}^M\, \xd{}\cdot\partial_{y}
-2\,\xd{}^M\, \yd{}\cdot\partial_{x} + (\xd{}\cdot\yd{}-R_1R_2)\, (\partial_{x^M}
-\partial_{y^M})\right ]\nn\\&&
\qquad\times \frac{1}{(x-z)^2\, (y-z)^2} \, .
\label{vecH}
\eea
After taking the derivatives in \eqn{vecH} and inserting the 
parameterization \eqn{parametrization} all angular integrals are elementary
and may be performed straightforwardly. Interestingly enough one 
finds the following identity
\be
\int_0^{2\pi}d\tau\, d\sigma\, \frac{\yd{}^M\, \xd{}\cdot(\yd{}-z)}
{(x-z)^2\,[(y-z)^2]^2} = 
\int_0^{2\pi}d\tau\, d\sigma\, \frac{\xd{}^M\, \yd{}\cdot(\xd{}-z)}
{[(x-z)^2]^2\,(y-z)^2} 
\ee 
which lets the fifth component of $H^M$ vanish and reduces \eqn{vecH} to
\be
H^\mu(z)= -2 \int_0^{2\pi}d\tau\, d\sigma\, \frac{\xd{}\cdot\yd{}-R_1R_2}
{(x-z)^2\,(y-z)^2}\, \left [ \,\frac{x^\mu-z^\mu}{(x-z)^2}-
\frac{y^\mu-z^\mu}{(y-z)^2}\, \right ] \, .
\label{Hdef}
\ee
Using this the full H-graph may be represented as
\be
\feyn{H}= \ft{g^6N(N^2-1)}{8}\,
\ft{1}{(4\pi^2)^5}\, \int d^4z\, d^4w\, 
\frac{H(z)\cdot H(w)}{(z-w)^2}
\label{FullH}
\ee
Performing the angular integrations in \eqn{Hdef} one finds
$H^\mu(r)=(r_1\, H^\rho, r_2\, H^\rho, H^3,H^4)$ where
\bea
H^\rho&=&
-\frac 2\rho\,\Bigl[\frac{4\,\pi^2\,\rho\,R_1\,
R_2\,(-R_1 + \sqrt{\rho^2 + r_3^2 + r_4^2})\,(R_1 + 
\sqrt{\rho^2 + r_3^2 + r_4^2})}{L_1^{3/2}\,\sqrt{L_2^h}} \nn\\&& 
- \frac{4\,\pi^2\,r\,
  R_1\,R_2\,(h^2 + \rho^2 - R_2^2 - 2\, h\,r_3 + r_3^2 + r_4^2)}{\sqrt{L_1}\,
 (L_2^h)^{3/2}}\nn\\&&
 + \frac{\pi^2\,(-\sqrt{L_2^h} + \rho^2 + R_2^2 + (h - r_3)^2 + 
            r_4^2}{2\,
              L_1^{3/2}\,\sqrt{L_2^h}\,
              \rho^3} \,\Bigl( -L_1^{3/2} + 8\,\rho^4\,R_1^2 
- 8\,\rho^2\,R_1^4 + R_1^6
\nn\\&&
 -  8\,\rho^2\,R_1^2\,(\rho^2 + r_3^2 + r_4^2) + 
                  3\,R_1^4\,(\rho^2 + r_3^2 + r_4^2) + 
                  3\,R_1^2\,(\rho^2 + r_3^2 + r_4^2)^2 + (\rho^2 + r_3^2 + r_4^2)^3
\Bigr )\nn\\&&
 + \frac{\pi^2\,(-1 + (\rho^2 + R_1^2 + r_3^2 + r_4^2)/\sqrt{L_1})}
{2\,(L_2^h)^{3/2}\,\rho^3}\,\Bigr(-h^6 + 
                  (L_2^h)^{3/2} - \rho^6 + 6\,h^5\,r_3 \nn\\&&
-   3\,\rho^4\,(R_2^2 + r_3^2 + r_4^2) - (R_2^2 + r_3^2 + r_4^2)^3
 -  3\,h^4\,(\rho^2 + R_2^2 + 5\,r_3^2 + r_4^2) \nn\\&&
+ 
                  4\,h^3\,r_3\,(3\,\rho^2 + 3\,R_2^2 + 5\,r_3^2 + 3\,r_4^2)
 + 
                  \rho^2\,(R_2^2 + r_3^2 + r_4^2)\,(5\,R_2^2 - 3\,(r_3^2 + r_4^2))
 \nn\\&& - 
                  h^2\,(3\,\rho^4 - 2\,\rho^2\,(R_2^2 - 9\,r_3^2 - 3\,r_4^2) + 
                        3\,(R_2^2 + r_3^2 + r_4^2)\,(R_2^2 + 5\,r_3^2 + r_4^2))
 \nn\\&& + 
                  2\, h\,r_3\,\Bigl\{3\,\rho^4 + 3\,(R_2^2 + r_3^2 + r_4^2)^2 + 
                        \rho^2\,(-2\,R_2^2 + 6\,(r_3^2 + r_4^2))\Bigr\} 
\Bigr)  \Bigr ]
\label{H1Final}
\eea
and
\bea
H^3&=&
-\frac{8\,\pi^2}{ L_1^{3/2}\,(L_2^h)^{3/2}}
\,\Bigl[L_1\,R_1\,R_2\,(h - r_3)\,(R_2^2 + (h - r_3)^2 + r_4^2 + \rho^2) 
\nn\\&&\quad+ 
   L_2^h\,R_1^2\,r_3\,(-\sqrt{L_2^h} + R_2^2
 + (h - r_3)^2 + r_4^2 + \rho^2) + 
   L_2^h\,R_1\,R_2\,r_3\,(R_1^2 + r_3^2 + r_4^2 + \rho^2) \nn\\&& \quad
+ 
   L_1\,R_2^2\,(h - r_3)\,(-\sqrt{L_1} + R_1^2 + r_3^2 + r_4^2 + \rho^2)\Bigr]
\label{H3Final}
\eea
and
\bea
H^4&=&
-\frac{8\,\pi^2\,r_4}{L_1^{3/2}\, (L_2^h)^{3/2}}
\,\Bigl[-(L_2^h)^{3/2}\,R_1^2 + h^2\,R_1\,(L_2^h\,R_1 - L_1\,R_2) - 
   2\, h\,R_1\,(L_2^h\,R_1 - L_1\,R_2)\,r_3 \nn\\&&\quad
+ L_2^h\,R_1\,(R_1 + R_2)\,
(R_1\,R_2 + r_3^2 + r_4^2 + 
     \rho^2) + L_1\,R_2\,(\sqrt{L_1}\,R_2 \nn\\&&\quad
- (R_1 + R_2)\,(R_1\,R_2 + 
r_3^2 + r_4^2 + \rho^2))\Bigr]
\label{H4Final}
\eea
with $\rho^2=r_1^2+r_2^2$.
Again this is as far as one gets analytically for the general geometry 
--  a number of further
integrals may be performed in \eqn{FullH} in the flat ($h=0$) or ``cake''
($R_1=R_2$) cases.

\subsection{Putting Everything Together}

We have thus shown for our geometry that, due to subtle cancellations,
the Maldacena-Wilson loop is completely finite to ${\cal O}(g^6)$.
Let us now verify whether or not the finite parts of
non-ladder diagrams cancel as well: We simply have to add 
the finite contributions of the interactive graphs worked out in sections
3.2 -3.5. For simplicity, as discussed above, 
we will content ourselves with the case $R:=R_1=R_2$, $h \neq 0$.
The quickest, yet safe way to arrive at the answer is to numerically
compute the contributions of all finite parts for some specific values
of $R$ and $h$ and to subsequently add the obtained numbers. This can be
done with great accuracy since our final expressions are finite,
low dimensional integrals. We tested various combinations of
$R$ and $h$ and found that the sum of finite parts yields in all instances
a non-zero number, exceeding the margin of error by many orders of
magnitude. In conclusion, to ${\cal O}(g^6)$ non-ladder diagrams
no longer cancel. Therefore no unknown ``non-renormalization theorem''
is at work, and we explicitly demonstrated that the quite amazing 
${\cal O}(g^4)$ cancellations observed
in the calculations of \cite{Erickson:2000af} are 
{\it not} generic. Our ultimate goal, however, is to verify whether
internal vertex diagrams contribute to the static potential in order
to explain the discrepancy between 
eqs.(\ref{rectangle}) and \eqn{ladderpotential}. We will now apply
our results to obtain some new insights into this question.

\section{The Static Potential Limit}

The geometric set-up analyzed in this work is clearly sufficiently rich to
recover the case of infinite, anti-parallel lines, corresponding
to the static potential limit: We simply take the radii of the circles
to infinity while keeping their separation finite. Luckily, in this limit the 
contributions of all graphs found above can be worked out explicitly.
We can therefore analytically demonstrate the non-cancellation of internal 
vertex diagrams, serving as a check on the numerical result of the
previous section. We will take the following limiting procedure:
\be
R_1=R_2=:R, \qquad T:=2 \pi R \rightarrow \infty \qquad {\rm while} \qquad 
L:=h \rightarrow 0
\label{limit1}
\ee  

Let us briefly discuss the subtle issue of the precise definition
of the static potential. For a rectangle, cf.~section 1, the measured
operator is a single loop and the usual definition is
$V(L)=-1/T \log \langle \W[C] \rangle$. Clearly this assumes the 
loop operator to be of the form $\W[C]=\exp [V(L)~T ]$. Explicit
calculations 
\cite{Erickson:2000qv},\cite{Erickson:2000af} demonstrate that
{\it perturbatively} this is not the case\footnote{E.g.~at 
${\cal O}(g^4)$ a term $\frac{T}{L} \log \frac{T}{L}$ appears. There is
some discussion in \cite{Erickson:2000qv},\cite{Erickson:2000af}
arguing that this term should be replaced by
$\frac{T}{L} \log \frac{1}{\lambda}$. It is not obvious to us whether
this is fully consistent, and in particular how to extend this procedure
to higher loops. }. In our case, we should analogously define the
static potential through 
$V(L)=-1/T \log \langle \W(C_1)\, \W(C_2) \rangle_{c}$.
The just mentioned problem of the perturbative definition of
$V(L)$ is further aggravated since now the connected two-loop correlator
does not even exponentiate in terms of a power series in the 
gauge coupling $g^2$.
We will therefore avoid any attempt to define $V(L)$ at weak coupling,
and use the notion ``static potential limit'' simply as a
{\it fa\c con de parler} to denote the perturbative evaluation of
$\langle \W[C] \rangle$ and $\langle \W(C_1)\, \W(C_2) \rangle_{c}$
in the geometric limit of infinite anti-parallel lines:
\be
\langle \W(C_1)\, \W(C_2) \rangle_{c} \Bigg|_{\ft TL \rightarrow \infty}=
\sum_{k=1}^{\infty}~W_{2 k}(T/L)~g^{2 k}
\label{expansion}
\ee
However, we do expect that at strong coupling 
$V(L)$ turns out to be well defined
-- at any rate, this is a prediction of AdS/CFT, cf.~eq.(\ref{rectangle}).

Let us now find the first three coefficients $W_{2 k}(T/L)$ in
eq.\eqn{expansion} from the results of section 3. 
Applying the limiting procedure \eqn{limit1},
$W_2$ and $W_4$ are immediately found from eqs.\eqn{1Gluon},\eqn{2Gluon}:
\be
W_2=\frac{g^2 N}{4 \pi}~\frac{T}{L}
\ee
and
\be
W_4=\frac{g^4 N^2}{32~\pi^2}\bigg(\frac{T}{L}\bigg)^2
-\frac{g^4 N^2}{32~\pi}  \frac{T}{L} 
\ee
To two-loop order the contributions of the ladder graphs are
directly obtained from the explicit results 
eqs.\eqn{3Gluon},(\ref{ladderfinal}),(\ref{ladderfinalnp}):
\bea
W_6^{\rm ladders}&=&\frac{g^6 N^3}{384~\pi^3}\bigg(\frac{T}{L}\bigg)^3
-\frac{g^6 N (N^2-1)}{32~\pi^4}~\bigg(\frac{T}{L}\bigg)^2
\Bigg(\, \log \left(4\pi\frac{T}{L}\right)+1\,\Bigg)
 \nn\\&& \nn\\&&\qquad + \frac{g^6 N (N^2-1)}{32~\pi^3}~\frac{T}{L}\,
\log \, \frac{T}{L}+
{\cal O}\Bigg(\frac{T}{L}\Bigg)
\label{W6ladders}
\eea
where we did not write out the subleading terms growing like 
$T/L$ and $\log T/L$.
 
Let us now investigate the contributions of the non-ladder diagrams
in the static potential limit. 
In the limit \eqn{limit1} the finite parts of the IY-graph integrals
\eqn{IY1Final} and \eqn{IY2Final} are dominated
by the region around $\rho\sim R$, $r_3\sim 0,h$ and $r_4\sim 0$.
The leading contribution in the static potential limit may then 
be extracted from the reduced integrals
\bea
\int (\mbox{IY})_{1}\Bigg |_{\ft T L\rightarrow \infty} &=&
-\frac{g^6N(N^2-1)}{128\pi^3}\, \frac{R}{h}\, 
 \int_0^\infty d\rho \int_{-\infty}^\infty dr_3\int_{-\infty}^\infty dr_4\nn\\
&&\quad\times
\frac{1}{(\rho-R)^2+r_3^2+r_4^2}\, \frac{1}{[\,(\rho-R)^2+(r_3-h)^2+r_4^2\,]^{1/2}}
\nn\\&&\nn\\
&=& -\frac{g^6N(N^2-1)}{64\pi^3}\, \frac{T}{L}\log \frac{T}{L} + 
{\cal O}\left(\frac {T}{L}\right)
\label{IY1Pot}
\eea
and
\bea
\int (\mbox{IY})_{2}\Bigg |_{\ft TL\rightarrow \infty} &=&
-\frac{g^6N(N^2-1)}{64\pi^3}\, 
 \int_0^\infty d\rho \int_{-\infty}^\infty dr_3\int_{-\infty}^\infty dr_4\nn\\
&&\quad\times
\frac{h-r_3}{[\, (\rho-R)^2+r_3^2+r_4^2\, ]^{1/2}}\, 
\frac{\log [\, 2 \,\frac{[\, (\rho-R)^2+r_3^2+r_4^2\, ]^{1/2}}{R}\,]}
{[\,(\rho-R)^2+(r_3-h)^2+r_4^2\,]^{3/2}}
\nn\\&&\nn\\
&=& \frac{g^6N(N^2-1)}{32\pi^2}\, \log \frac{T}{L} + {\cal O}(1) \,.
\label{IY2Pot}
\eea
telling us that this graph is dominated by the $(\mbox{IY})_{1}$ 
piece in the $T/L\rightarrow\infty$ limit. The mirror
IY${}^T$-graph yields the same contributions \eqn{IY1Pot} and \eqn{IY2Pot}
to the static potential limit.
Turning to the X-graph in the limit \eqn{limit1}
one finds from \eqn{XFinal} the leading behaviour
\bea
\feyn{X}\Bigg |_{\ft TL\rightarrow \infty} &=&\frac{g^6N(N^2-1)\, R}{256\pi^3}\, 
 \int_0^\infty d\rho \int_{-\infty}^\infty dr_3\int_{-\infty}^\infty dr_4\nn\\
&&\quad\times
\frac{1}{(\rho-R)^2+r_3^2+r_4^2}\, \frac{1}{(\rho-R)^2+(r_3-h)^2+r_4^2}
\nn\\&&\nn\\
&=& \frac{g^6N(N^2-1)}{512\pi}\, \frac{T}{L} + {\cal O}\left(\log \frac {T}{L}\right
) \, .
\eea
In order to extract the static potential limit of the H-graph it is 
technically preferable to consider a different limit than \eqn{limit1}
and rather take  
\be
h=0, \qquad T:=2 \pi R_2 \rightarrow \infty \qquad {\rm while} \qquad
L:=|R_2-R_1| \rightarrow 0 \, ,
\label{limit2}
\ee
which leads to rotational symmetry also in the 3-4 plane. For the leading
contribution in $T/L$ this limit is completely equivalent to \eqn{limit1}.
Introducing $\sigma$ as the radius in the 3-4 plane, 
the integral \eqn{FullH} reduces in this flat geometry to
\bea
\feyn{H}\Bigg |_{h=0} &=&\ft{g^6N(N^2-1)}{8\cdot (4\pi^2)^4}
\int_0^\infty d\rho \, d\bar\rho\, d\sigma\,
d\bar\sigma \int_0^{2\pi} d \phi\,  d \theta \, \rho\,\bar\rho\,\sigma\,
\bar\sigma \nn\\&&\quad\times
\frac{H^\rho\, H^{\bar\rho}\,\rho\,\bar\rho \,\cos \phi 
+H^\sigma\, H^{\bar\sigma}\,\sigma\bar\sigma\, \cos \theta}
{\rho^2+\bar\rho^2+\sigma^2+\bar\sigma^2- 2\rho\,\bar\rho\,\cos\phi
-2 \sigma\,\bar\sigma\, \cos\theta}
\label{elvis}
\eea
where the barred and unbarred quantities are associated with the
1-2 and 3-4 plane radii of the
space-points $w$ and $z$ of \eqn{FullH} respectively, and
$\phi$ and $\theta$ are the relative angles. In the limit \eqn{limit2}
this integral is dominated by the region $(\rho,\bar\rho)\sim [R_1,R_2]$,
$(\sigma,\bar\sigma)\sim 0$, $\phi\sim 0$ and $\theta\in [0,2\pi]$.
Within this region the functions $H^\rho$ and $H^\sigma$ following from
\eqn{H1Final},\eqn{H3Final},\eqn{H4Final} read
\bea
H^\rho&=& 2\pi^2\,\frac{R_1-R_2}{R_1+R_2}\, \frac{(\rho-R_1)\, (\rho-R_2)}
{[\,(\rho-R_1)^2+\sigma^2\,]^{3/2}\,  [\,(\rho-R_2)^2+\sigma^2\,]^{3/2}}\nn\\
H^\sigma&=&2\pi^2\, (R_1-R_2)\, \frac{\rho-(R_1+R_2)/2}
{[\,(\rho-R_1)^2+\sigma^2\,]^{3/2}\,  [\,(\rho-R_2)^2+\sigma^2\,]^{3/2}}
\label{HSPL}
\eea
Plugging \eqn{HSPL} into the static potential limit of \eqn{elvis}
and performing the integral over $\theta$
one discovers that the dependence on $T/L$ may be scaled out of the 
integral. Finally one has
\be
\feyn{H}\Bigg|_{\ft TL\rightarrow \infty} = c\cdot g^6N(N^2-1)\,
\frac{T}{L}
+{\cal O}\left(\log \frac{T}{L}\right ) \qquad
\qquad c=7.23(9)\cdot 10^{-5}
\ee
where $c$ is a number expressed through a finite yet complicated 
five dimensional integral independent of $T/L$, which we evaluated
numerically.

To summarize, we have thus analytically proven the non-cancellation
of the internal vertex diagrams. In particular, we notice that the
$T/L$ behavior is different for the various graphs: Independently of
the symmetry factor they cannot possibly cancel. The 
strongest contribution in the non-ladder sector comes from the IY-graph
so that
\be
W_6^{\rm non-ladders}= - \frac{g^6 N (N^2-1)}{32~\pi^3}~\frac{T}{L}\,
\log \, \frac{T}{L}+
{\cal O}\Bigg(\frac{T}{L}\Bigg)
\label{W6non-ladders}
\ee
which presicely cancels the subleading $T/L\, \log T/L$ ladder-term in
\eqn{W6ladders}.

Despite the fact that the non-ladder-diagrams do not
cancel, our results indicate that their contribution to the limit of
infinite anti-parallel lines is rather weak 
(at ${\cal O}(g^6)$ $T/L\,\log T/L$ vs.~$(T/L)^3$).  Unfortunately their
quantitative influence on the strong coupling potential is difficult
to estimate.
Naively, one could speculate that the fact that the vertex contributions are 
strongly subleading might explain the near coincidence of 
eqs.(\ref{rectangle}),(\ref{ladderpotential}).
However the issue is rather subtle: In the planar ($N =\infty$) limit
we expect that summing up perturbation theory correctly reproduces the
non-perturbative physics\footnote{There is a caveat:
A large $N$ phase transition, triggered by the increase in the
number of non-planar diagrams, might take place at some finite
value of the `t Hooft coupling $\lambda = N g^2$: In that case the
weak and strong coupling phase might turn out to be analytically unrelated,
yielding one possible explanation for the discrepancy between
the strong coupling ladder-approximation eq.\eqn{ladderpotential} 
and the AdS/CFT result eq.\eqn{rectangle}.}. In principle it is therefore
conceivable that the ladder-approximation still captures
the correct gauge-theory result in the large $N$, strong-coupling and 
$T/L\rightarrow \infty$ limits. If true, and if eq.\eqn{ladderpotential}
is correct this would contradict the AdS/CFT prediction eq.\eqn{rectangle}. 
Clearly we need to increase our
understanding on how the subleading terms in the $T/L$ expansion of
the coefficients $W_{2 k}$ influence the strong coupling static
potential extracted from eq.\eqn{expansion}. This will be left for
future work.

\section{Outlook}

The AdS/CFT correspondence has been inspiring since it
appears to allow for analytic calculations in strongly coupled $U(N)$ gauge 
theory, at least when $N$ is infinite and ${\cal N}$ (the number of
supersymmetries) is large (${\cal N}=4$).
However, we feel more attempts should be made to state the
consequences of the correspondence for the gauge theory more precisely
in order to perform quantitative, non-kinematical analytic tests. 
A perfect comparison would be to use supersymmetry to
obtain a non-trivial strong coupling result on the gauge theory
side that unequivocally agrees with the ``corresponding'' 
classical supergravity computation. Clearly non-unitary
Maldacena-Wilson loops, as opposed to ordinary unitary Wilson loops, are
perfect candidates for such a test: They appear to be completely 
{\it finite gauge invariant} observables with intriguing properties
on both sides of the correspondence. As such, they deserve to be much
more carefully studied: 

\noindent$\bullet$ To date, no rigorous proof exists in ${\cal N}=4$
gauge theory that all perturbative corrections cancel in the case of a 
straight line:
\be
\langle \W[C] \rangle =1, \qquad  C={\rm straight} \; {\rm line}.
\ee  

\noindent$\bullet$ It has been argued in \cite{Drukker:2000rr}
but by no means proven that only ladder diagrams contribute
to a circular Maldacena-Wilson loop for all finite gauge groups
$U(N)$ and $SU(N)$ (see e.g.~eq.(\ref{su2})).
If true this would lead to
\be 
\langle \W[C] \rangle ={1 \over N} L^1_{N-1}(-g^2/4)~e^{g^2/8},
\qquad C={\rm circle},
\ee
where $L^1_{N-1}$ is a Laguerre polynomial of degree $N$ in $g^2$.
In \cite{Akemann:2001st} a variation of the anomaly arguments of
\cite{Drukker:2000rr} proposing a zero-dimensional matrix model
description of the circular loop was considered. However, the relation
of the considered matrix model potential to the full field theoretic
problem remains unclear.
An important further question is the strength of instanton contributions to
the Maldacena-Wilson loop -- preliminary results were reported in
\cite{Bianchi:2001jg}.

\noindent$\bullet$  The subtle cancellations of divergences found
in this and previous work required a serious calculational effort;
they are far from obvious.
An all-orders result for arbitrary smooth contours would
be very desirable.
In fact one should prove the highly non-trivial property
\be
0< \left | \langle \W[C] \rangle  \right | < \infty \qquad C
= ({\rm any?}) \; {\rm contour}.
\ee

\noindent$\bullet$ A perfect quantitative test for the correspondence
would be the verification of the strong coupling static potential
as predicted by AdS/CFT: How to find in the gauge theory the non-trivial
number in front of the Coulomb potential (see eq.(\ref{rectangle}))
\be
V(L) \sim 4 \, \pi^2 \, \Gamma^{-4} (1/4) \; L^{-1} \; ? 
\ee
In the present work we discussed some of the subtleties surrounding this
major challenge. Our result shows that non-ladder diagrams 
are non-vanishing but subleading,
and one wonders whether the ladder approximation of
\cite{Erickson:2000qv} could be improved in order to systematically
approximate the correct result.  

\noindent$\bullet$ Our results can also be used to consider the limit
$R_2\rightarrow 0$ and $h \gg R_1$, which should be related to
the scenario of \cite{Semenoff:2001xp}.

\noindent$\bullet$ There has been some confusion in the literature
as to whether the Maldacena-Wilson loop operator (\ref{maldaloop})
is the final object to be compared to supergravity. The issue is
whether or not supersymmetry requires also fermionic degrees on
the boundary.   

\noindent$\bullet$ The existing perturbative studies do not make
any explicit use of supersymmetry. 
In fact, in this work
supersymmetry only entered in the fermionic contribution to
the gauge/scalar self energy, and, quite indirectly, in the
``BPS'' property of the Maldacena-Wilson loop \cite{Drukker:1999zq}.
Is there a more appropriate
gauge (e.g.~Mandelstam-Leibbrandt) or a better technique 
(e.g.~supergraphs) that renders higher order calculations feasible?

\section*{Acknowledgments.} 
The authors would like to thank G.~Arutyunov, A.~Hartl, S.~Kuzenko,
H.~Nicolai, Y.~Schr\"oder, G.~Semenoff, S.~Silva and S.~Theisen for
useful discussions.


\begin{thebibliography}{1}


%\cite{'tHooft:1974jz}
\bibitem{'tHooft:1974jz}
G.~'t Hooft,
``A Planar Diagram Theory For Strong Interactions,''
Nucl.\ Phys.\ B {\bf 72} (1974) 461.
%%CITATION = NUPHA,B72,461;%%

%\cite{Maldacena:1998re}
\bibitem{Maldacena:1998re}
J.~Maldacena,
``The large N limit of superconformal field theories and supergravity,''
Adv.\ Theor.\ Math.\ Phys.\  {\bf 2} (1998) 231
[Int.\ J.\ Theor.\ Phys.\  {\bf 38} (1998) 1113]
[hep-th/9711200].
%%CITATION = HEP-TH 9711200;%%

%\cite{Maldacena:1998im}
\bibitem{Maldacena:1998im}
J.~Maldacena,
``Wilson loops in large N field theories,''
Phys.\ Rev.\ Lett.\  {\bf 80} (1998) 4859
[hep-th/9803002].
%%CITATION = HEP-TH 9803002;%%

%\cite{Berenstein:1999ij}
\bibitem{Berenstein:1999ij}
D.~Berenstein, R.~Corrado, W.~Fischler and J.~Maldacena,
``The operator product expansion for Wilson loops and surfaces in the  large N limit,''
Phys.\ Rev.\ D {\bf 59} (1999) 105023
[hep-th/9809188].
%%CITATION = HEP-TH 9809188;%%

%\cite{Aharony:2000ti}
\bibitem{Aharony:2000ti}
O.~Aharony, S.~S.~Gubser, J.~Maldacena, H.~Ooguri and Y.~Oz,
``Large N field theories, string theory and gravity,''
Phys.\ Rept.\  {\bf 323} (2000) 183
[hep-th/9905111].
%%CITATION = HEP-TH 9905111;%%

%\cite{Erickson:2000qv}
\bibitem{Erickson:2000qv}
J.~K.~Erickson, G.~W.~Semenoff, R.~J.~Szabo and K.~Zarembo,
``Static potential in N = 4 supersymmetric Yang-Mills theory,''
Phys.\ Rev.\ D {\bf 61} (2000) 105006
[hep-th/9911088].
%%CITATION = HEP-TH 9911088;%%

%\cite{Erickson:2000af}
\bibitem{Erickson:2000af}
J.~K.~Erickson, G.~W.~Semenoff and K.~Zarembo,
``Wilson loops in N = 4 supersymmetric Yang-Mills theory,''
Nucl.\ Phys.\ B {\bf 582} (2000) 155
[hep-th/0003055].
%%CITATION = HEP-TH 0003055;%%

%\cite{Drukker:2000rr}
\bibitem{Drukker:2000rr}
N.~Drukker and D.~J.~Gross,
``An exact prediction of N = 4 SUSYM theory for string theory,''
hep-th/0010274.
%%CITATION = HEP-TH 0010274;%%

%\cite{Zarembo:1999bu}
\bibitem{Zarembo:1999bu}
K.~Zarembo,
``Wilson loop correlator in the AdS/CFT correspondence,''
Phys.\ Lett.\ B {\bf 459} (1999) 527
[hep-th/9904149];\\
P.~Olesen and K.~Zarembo,
``Phase transition in Wilson loop correlator from AdS/CFT correspondence,''
hep-th/0009210.
%%CITATION = HEP-TH 0009210;%%

%\cite{Zarembo:2001jp}
\bibitem{Zarembo:2001jp}
K.~Zarembo,
``String breaking from ladder diagrams in SYM theory,''
JHEP {\bf 0103} (2001) 042
[hep-th/0103058].
%%CITATION = HEP-TH 0103058;%%

%\cite{Drukker:1999zq}
\bibitem{Drukker:1999zq}
N.~Drukker, D.~J.~Gross and H.~Ooguri,
``Wilson loops and minimal surfaces,''
Phys.\ Rev.\ D {\bf 60} (1999) 125006
[hep-th/9904191].
%%CITATION = HEP-TH 9904191;%%

%\cite{Akemann:2001st}
\bibitem{Akemann:2001st}
G.~Akemann and P.~H.~Damgaard,
``Wilson loops in N = 4 supersymmetric Yang-Mills theory from random  matrix theory,''
Phys.\ Lett.\ B {\bf 513} (2001) 179
[hep-th/0101225].
%%CITATION = HEP-TH 0101225;%%

%\cite{Bianchi:2001jg}
\bibitem{Bianchi:2001jg}
M.~Bianchi, M.~B.~Green and S.~Kovacs,
``Instantons and BPS Wilson loops,''
hep-th/0107028.
%%CITATION = HEP-TH 0107028;%%

%\cite{Semenoff:2001xp}
\bibitem{Semenoff:2001xp}
G.~W.~Semenoff and K.~Zarembo,
``More exact predictions of SUSYM for string theory,''
hep-th/0106015.
%%CITATION = HEP-TH 0106015;%%

\end{thebibliography}
\end{document}